\documentclass[aps,prb,10pt,twocolumn,a4paper,superscriptaddress,nofootinbib,longbibliography]{revtex4-2}

\usepackage[colorlinks = true,
            linkcolor = blue,
            urlcolor  = blue,
            citecolor = blue,
            anchorcolor = blue,
            unicode]{hyperref}

\usepackage{graphicx}
\usepackage{amsmath}
\usepackage{amssymb}
\usepackage{amsfonts}
\usepackage{color}

\DeclareMathOperator{\maxmin}{max(min)}

\begin{document}

\title{Influence of capacitance and thermal fluctuations on the Josephson diode effect in asymmetric higher-harmonic SQUIDs}

\author{G.~S.\ Seleznev}
\affiliation{L.~D.\ Landau Institute for Theoretical Physics RAS, 142432 Chernogolovka, Russia}
\affiliation{Moscow Institute of Physics and Technology, 141700 Dolgoprudny, Russia}

\author{Ya.~V.\ Fominov}
\affiliation{L.~D.\ Landau Institute for Theoretical Physics RAS, 142432 Chernogolovka, Russia}
\affiliation{Moscow Institute of Physics and Technology, 141700 Dolgoprudny, Russia}
\affiliation{Laboratory for Condensed Matter Physics, HSE University, 101000 Moscow, Russia}

\begin{abstract}
Asymmetric two-junction SQUIDs with different current-phase relations in the two Josephson junctions, involving higher Josephson harmonics, demonstrate a flux-tunable Josephson diode effect (asymmetry between currents flowing in the opposite directions, which can be tuned by the magnetic flux through the interferometer loop).
We theoretically investigate influence of junction capacitance and thermal fluctuations on performance of such Josephson diodes.
Our main focus is on the ``minimal model'' with one junction in the SQUID loop possessing the sinusoidal current-phase relation and the other one featuring additional second harmonic.
Capacitance generally weakens the diode effect in the resistive branch (R state) of the current-voltage characteristic (CVC) both in the absence and in the presence of external ac irradiation.
At the same time, it leads to qualitatively new features of the Josephson diode effect such as asymmetry of the retrapping currents (which are a manifestation of hysteretic CVC).
In particular, the limiting case of the single-sided hysteresis becomes accessible. In its turn, thermal fluctuations are known to lead to nonzero average voltages at any finite current, even below the critical value. We demonstrate that in the diode regime, the fluctuation-induced voltage can become strongly (exponentially) asymmetric. In addition, we find asymmetry of the switching currents arising both due to thermal activation and due to Josephson plasma resonances in the presence of ac irradiation.
\end{abstract}

\date{25 September 2024}

\maketitle

\tableofcontents

\section{Introduction}
\label{sec:intro}


While superconducting systems demonstrating nonreciprocal transport properties (the diode effect) have been known for a long time \cite{KulikYansonBook, BaroneBook, Moll2023}, they have become the focus of many studies during recent years.
This superconducting diode effect (SDE) is currently actively investigated both theoretically and experimentally in various physical systems \cite{Nadeem2023}. The physical mechanisms causing the SDE turn out to be quite diverse, so it can be considered as a spectacular manifestation of various fundamental physical processes. At the same time, the SDE can potentially find useful applications in superconducting electronic devices.

The necessary ingredients of the SDE are usually broken time-reversal and inversion symmetries, which can be realized, e.g., due to magnetic field (or exchange field in ferromagnets) and spin-orbit coupling (or spatial asymmetry). The SDE can also be realized due to vortices moving in asymmetric potentials or due to current-generated magnetic fields.
The above mechanisms have been theoretically studied and experimentally demonstrated in many publications
\cite{Levitov1985, Edelstein1996, Majer2003.PhysRevLett.90.056802, Villegas2003, Vodolazov2005.PhysRevB.72.064509, deSouzaSilva2006, Morelle2006, Aladyshkin2010.10.1063/1.3474622, Silaev2014, Wakatsuki2017, Yasuda2019,Ando2020,Lyu2021, Daido2022.PhysRevLett.128.037001, Yuan2022, Ilic2022.PhysRevLett.128.177001, He2022, Kokkeler2022.PhysRevB.106.214504, Karabassov2022.PhysRevB.106.224509, Suri2022, Levichev2023.PhysRevB.108.094517, Hasan2024.PhysRevB.110.024508}.

Similar physical mechanisms
\cite{Krasnov1997.PhysRevB.55.14486, Yokoyama2014.PhysRevB.89.195407, Chen2018.PhysRevB.98.075430, Kopasov2021.PhysRevB.103.144520,Golod2022, Baumgartner2022, Halterman2022.PhysRevB.105.104508, Pal2022, Zhang2022.PhysRevX.12.041013, Davydova2022, Kokkeler2024.10.21468/SciPostPhys.16.2.055, Debnath2024.PhysRevB.109.174511, Chatterjee2024, Meyer2024.10.1063/5.0211491, Zhang2022arXiv, Sivakumar2024arXiv, Karabassov2024arXiv}
can lead to the SDE in various types of Josephson junctions (JJs); in this context it is called the Josephson diode effect (JDE).
This brings the rich physics of the Josephson effect \cite{KulikYansonBook, BaroneBook, LikharevBook} into play. Asymmetry of the Josephson effect characteristics with respect to the current direction implies realization of the JDE.

Of particular interest are SQUIDs, tunable Josephson systems of interferometer type \cite{KulikYansonBook, BaroneBook, LikharevBook}. A basic system of this type is shown in Fig.~\ref{fig:SQUID}; the interferometer loop contains two JJs and is threaded by external magnetic flux $\Phi$. The up-down asymmetry of such a system (asymmetry between junctions $a$ and $b$) in the presence of magnetic flux may lead to the left-right asymmetry for the current (the JDE). This is exemplified by SQUIDs with asymmetry of effective inductances included into the two interferometer arms \cite{Fulton1972.PhysRevB.6.855, Peterson1979, BaroneBook}. While this effect has been known for a long time, miniaturization of SQUID systems diminishes inductive effects and thus suppresses this kind of the diode effect.

At the same time, it was recently demonstrated \cite{MikhailovMSThesis2020, Fominov2022.PhysRevB.106.134514, Souto2022.PhysRevLett.129.267702} that the up-down asymmetry of the SQUID due to higher Josephson harmonics in the current-phase relation (CPR) of the JJs can also lead to the JDE. This is so even in the absence of inductive effects (hence, the mechanism is effective even in the case of small systems). Generally, the JDE then takes place in the case of different harmonic content of the CPRs $I_a(\varphi)$ and $I_b(\varphi)$ of the two JJs.
The higher Josephson harmonics (contributions to the supercurrent of the form $\sin n\varphi$ with $n>1$, where $\varphi$ is the superconducting phase difference across a JJ) naturally arise in various types of JJs with not too low transparencies of their weak-link regions (represented by insulators, normal metals, or ferromagnets) \cite{KulikYansonBook, LikharevBook, Golubov2004review}.
Josephson elements with essential contribution of higher harmonics can also be engineered on purpose \cite{Messelot2024arXiv, Leblanc2024arXiv}.

The JDE in the above-mentioned SQUIDs is absent only in certain special cases, e.g.,
(a) in a symmetric SQUID with $I_a(\varphi)=I_b(\varphi)$ (at arbitrary number and amplitudes of the harmonics),
(b) in the case when $I_a(\varphi)$ and $I_b(\varphi)$ are both described by the same single harmonic (with arbitrary amplitudes in the two junctions),
(c) at `trivial' values of the magnetic flux $\Phi$ (integer of half-integer in units of the flux quantum $\Phi_0$).
Otherwise, the JDE is generally present. The ``minimal model'' of the asymmetric higher-harmonic SQUID is the case in which one JJ has the sinusoidal CPR and the other one features additional second Josephson harmonic \cite{MikhailovMSThesis2020,Fominov2022.PhysRevB.106.134514},
\begin{equation} \label{eq:minmod}
I_a(\varphi) = I_{a1} \sin\varphi,
\quad
I_b(\varphi) = I_{b1} \sin\varphi + I_{b2} \sin 2\varphi.
\end{equation}

Various SQUIDs and SQUID-like systems effectively implementing the higher-harmonic JDE mechanism have already been investigated both theoretically and experimentally
\cite{Chen2018.PhysRevB.98.075430, Gupta2023,Greco2023.10.1063/5.0165259, Ciaccia2023.PhysRevResearch.5.033131, Bozkurt2023.10.21468/SciPostPhys.15.5.204, Ciaccia2024, Valentini2024, Zhang2024.10.21468/SciPostPhys.16.1.030, Souto2024.PhysRevResearch.6.L022002, Haenel2022arXiv, Li2023arXiv, Leblanc2023arXiv}.
The basic quantities of interest here are the direction-dependent critical currents ($I_c^+$ and $I_c^-$) and asymmetry of the current-voltage characteristic (CVC) $I(V)$ both in the absence and in the presence of external irradiation.
The CVC in Josephson systems can be described with the help of the standard resistively shunted junction (RSJ) model \cite{BaroneBook,LikharevBook}. The JDE in SQUIDs with higher harmonics is fully captured by this model once the proper CPR is plugged into it.

A natural extension of the RSJ model is the resistively and capacitively shunted junction (RCSJ) model which takes possible capacitance (charging) effects into account \cite{BaroneBook,LikharevBook}. While the mechanical analogy of the RSJ model corresponds to strongly damped motion, the RCSJ model adds inertial effects to it. As a result, it is possible to trace crossover between overdamped and underdamped regimes. A natural question then is how the presence of capacitance influences the JDE. The RSJ model can also be extended to include a fluctuating current to describe thermal fluctuations \cite{Ivanchenko1968, Ambegaokar1969.PhysRevLett.22.1364, Buttiker1983.PhysRevB.28.1268}, and one may expect that fluctuations lead to strong asymmetry of the CVC under the conditions of the JDE. In the context of the JDE, various charging and temperature effects have been studied before both theoretically and experimentally in Refs.\ \cite{Misaki2021.PhysRevB.103.245302, Wu2022, Steiner2023.PhysRevLett.130.177002, Haenel2022arXiv}.

In this paper, we analyze the influence of capacitance on the JDE in asymmetric higher-harmonic SQUIDs in different regimes, from underdamped to overdamped. We also consider asymmetries of the current caused by thermal fluctuations.

The paper is organized as follows:
In Sec.~\ref{sec:general}, we formulate general equations of the RCSJ model suitable for describing the SQUIDs with higher Josephson harmonics and underline basic features of the system which are essential for further analysis.
In Sec.~\ref{sec:CVC}, we analyze the main features of the asymmetric CVC of the minimal model with nonzero capacitance.
In Sec.~\ref{sec:irradiation}, we analyze the influence of capacitance on the CVC in the presence of external irradiation.
In Sec.~\ref{sec:Tfluct}, we consider manifestations of thermal fluctuations in the context of the JDE.
In Sec.~\ref{sec:discussion}, we discuss the obtained results and their possible applications.
In Sec.~\ref{sec:conclusions}, we present our conclusions.
Finally, some details of calculations are presented in the Appendices.

\section{Model and general equations}
\label{sec:general}

\subsection{Asymmetric SQUID}
\label{sec:SQUID}

\begin{figure}[t]
 \includegraphics[width=\columnwidth]{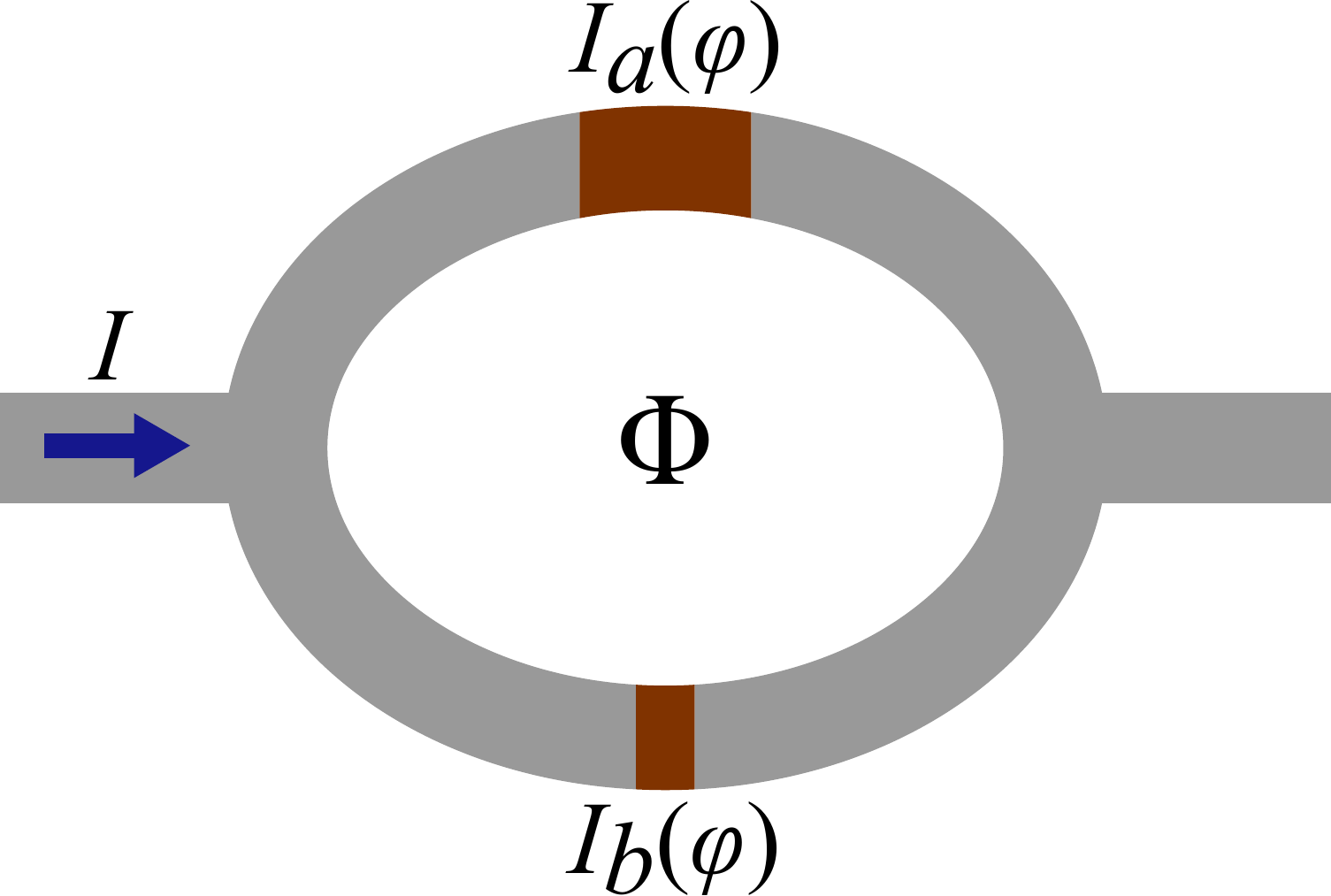}
 \caption{Asymmetric SQUID with different CPRs $I_a(\varphi)$ and $I_b(\varphi)$ in the two JJs.}
 \label{fig:SQUID}
\end{figure}

In this section, we present the theoretical model in which we investigate the JDE. It is an extension of the model described in Refs.\ \cite{MikhailovMSThesis2020, Fominov2022.PhysRevB.106.134514, Souto2022.PhysRevLett.129.267702}. We consider a two-junction asymmetric SQUID consisting of two JJs connected in parallel and possessing different CPRs with higher Josephson harmonics, see Fig.~\ref{fig:SQUID}.

We mainly focus on the minimal model in this paper. In this model, one junction has the standard sinusoidal CPR while the other one also has the second Josephson harmonic in its CPR, see Eq.\ \eqref{eq:minmod}.
The external magnetic field creates flux $\Phi$ through the SQUID loop. The flux leads to the difference between the phase jumps at the two JJs:
\begin{equation} \label{eq:phi}
  \varphi_a-\varphi_b = \phi, \qquad
  \phi = 2\pi \Phi/\Phi_0.
\end{equation}
Defining $\varphi$ as the average of the two phase jumps, we can write the effective CPR of the SQUID as
\begin{equation} \label{eq:Is}
I_{s}(\varphi) = I_{a} (\varphi + \phi/2) + I_{b}(\varphi - \phi/2).
\end{equation}
In the case of the minimal model, it takes the simple form
\begin{equation} \label{eq:CPR}
J(\varphi) = I_{s}(\varphi)/I_{1}(\phi) = \sin\varphi + A \sin(2 \varphi - \tilde{\phi}),
\end{equation}
where we define the amplitude of the first Josephson harmonic of the SQUID $I_{1}(\varphi)$, dimensionless amplitude of the second Josephson harmonic $A$, and  phase shift $\tilde{\phi}$ as
\begin{gather}
I_1 (\phi) = \sqrt{I_{a1}^2 + I_{b1}^2 +2 I_{a1} I_{b1} \cos\phi}, \quad A = I_{b2}/I_{1}(\phi),
\label{eq:I_1(phi)}
\\
\tan \gamma = \frac{I_{a1}-I_{b1}}{I_{a1}+I_{b1}} \tan\frac{\phi}{2}, \quad \tilde{\phi} = \phi + 2 \gamma(\phi).
\label{eq:beta}
\end{gather}
Equation \eqref{eq:CPR} describes the CPR of the whole SQUID as a single effective JJ.

\subsection{RCSJ model}

We are interested in asymmetries in the SQUID behavior (CVC, Shapiro steps, etc.) when the system is subject to external currents, dc current with amplitude  $I_{\mathrm{dc}}$ and ac current with amplitude $I_{\mathrm{ac}}$ and frequency $\Omega$. The extension of our model as compared to the model of Refs.\ \cite{MikhailovMSThesis2020,Fominov2022.PhysRevB.106.134514} is that we now consider the cases of nonzero capacitance $C$ and nonzero temperature $T$. We figure out how their presence affects the strength and manifestations of the JDE in the system. 

To describe the dynamics of our system, we use the RCSJ model. In this model, the Josephson equations take the following form:
\begin{gather}
\frac{\hbar C}{2e} \ddot{\varphi} + \frac{\hbar}{2eR} \dot{\varphi} + I_{s}(\varphi) 
=
I_{\mathrm{dc}} + I_{\mathrm{ac}}\cos(\Omega t +\delta) + I_{f}(t), 
\label{eq:BegJ1} \\
\label{eq:BegJ2}
V = (\hbar/2e) \dot{\varphi},
\end{gather}
where $R$ is the normal resistance, $V$ is the voltage bias across the SQUID, $I_f(t)$ is thermally induced fluctuating current, and $\delta$ is the initial phase of the ac current. 

We rewrite these equations in dimensionless variables. It can be done in several ways. The first form that we call $\beta$ representation is convenient for analysis of the CVC in the nonstationary (resistive) regime and in the small-capacitance limit. In this representation, time is measured in units of the oscillation time in the R state. As a result, the McCumber parameter $\beta$ appears:
\begin{equation}
    \beta = (2 e/\hbar) I_{1} R^2 C, \quad \tau = \omega_{J} t, \quad \omega_{J} = (2e/\hbar) I_{1} R, 
\end{equation}
where $\omega_{J}$ is the Josephson frequency. The McCumber parameter determines the strength of the charging effects (the larger this parameter is, the stronger the capacitive effects are). Equations \eqref{eq:BegJ1} and \eqref{eq:BegJ2} in this representation take the following form:
\begin{gather} \label{eq:J1}
\beta \frac{{d^2 \varphi}}{d \tau^2} + \frac{d \varphi}{d \tau} + J(\varphi) =  j_{\mathrm{dc}} + j_{\mathrm{ac}}\cos(\omega \tau +\delta) +\xi(\tau), \\
\label{eq:J2}
v = d \varphi/d \tau,
\end{gather}
where $j_{\mathrm{dc}/\mathrm{ac}} = I_{\mathrm{dc/ac}}/I_{1}$, $v = V/I_{1} R$, and $\omega = \Omega/\omega_{J}$. Thermal fluctuations of the current in Eq.\ \eqref{eq:J1} are considered as a white noise with correlator  $\langle \xi(\tau)\xi(\tau') \rangle = 2 \theta \delta(\tau - \tau')$, where $\theta = 2e T/\hbar I_\mathrm{1} = T/E_{J}$ is dimensionless temperature and $E_{J} = \hbar I_{1}/2e$ is the Josephson energy.

The second representation, which we call $\varepsilon$ representation, is more convenient for considering the junction behavior when energy dissipation in the system is small and for the analysis of the oscillations in the stationary (S) state. In this case, time is measured in units of the oscillation time of the particle in the potential well (see Sec.~\ref{sec:U}). In this representation, the new parameter appears in the Josephson equations:
\begin{equation}
\varepsilon = 1/\sqrt{\beta}, \quad \tilde{\tau} = \omega_{p} t, \quad \omega_{p} = \sqrt{2 e I_{1} / \hbar C},
\end{equation}
where $\omega_{p}$ is the plasma frequency and $\varepsilon$ has the meaning of the dissipation factor. In the $\varepsilon$ representation, Eqs.\ \eqref{eq:BegJ1} and \eqref{eq:BegJ2} take the following form:
\begin{gather} \label{eq:J1 epsilon}
\frac{d^2 \varphi}{d \tilde{\tau}^2} + \varepsilon \frac{d \varphi}{d \tilde{\tau}} + J(\varphi) =  j_{\mathrm{dc}} + j_{\mathrm{ac}}\cos(\tilde{\omega} \tilde{\tau} +\delta) +\xi(\tilde{\tau}), \\
\label{eq:J2 epsilon}
v =\varepsilon d \varphi/ d \tilde{\tau},
\end{gather}
where $\tilde{\omega} = \Omega / \omega_{p}$ and the noise correlator is $\langle \xi(\tilde{\tau})\xi(\tilde{\tau}') \rangle = 2 \theta \varepsilon \delta(\tilde{\tau} - \tilde{\tau}')$.
Equations \eqref{eq:J1}, \eqref{eq:J2}, and \eqref{eq:CPR} [or equivalently Eqs. \eqref{eq:J1 epsilon}, \eqref{eq:J2 epsilon}, and \eqref{eq:CPR}]  fully determine the CVC of the system, that is the dependence of the average voltage on the dc current,  $\langle \overline{v}(j_{\mathrm{dc}}) \rangle$, where $\overline{\dots \vphantom{v}}$ means time averaging and $\langle \dots \rangle$ means averaging over thermal fluctuations. 

Finally, we underline that in this paper we consider the system in the current-source regime with $j_{\mathrm{dc}} = \mathrm{const}$ and $j_{\mathrm{ac}} = \mathrm{const}$.

\subsection{Asymmetric potential}
\label{sec:U}

\begin{figure}[t]
\centering
 \includegraphics[width=0.95\columnwidth]{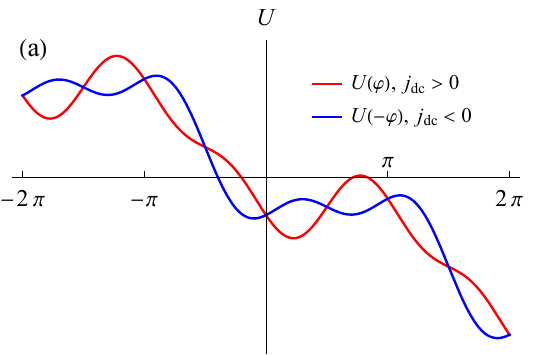}
 \hfill
 \includegraphics[width=0.95\columnwidth]{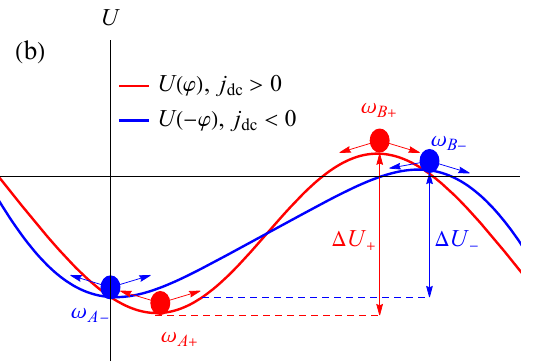}
\caption{(a) Potential $U(\varphi)$ from Eq.\ \eqref{eq:U(phi)} for $A = 1$, $\tilde{\phi} = \pi/2$, and $j_{\mathrm{dc}} = \pm 0.5$. The overall potential shape and particularly the number of minima depend on the current direction. (b)~Asymmetry of the main features of $U(\varphi)$: oscillation frequencies at the well bottom $\omega_{A \pm}$, curvatures of the barrier $\omega_{B \pm}$, and heights of the potential barriers $\Delta U_{\pm}$.}
 \label{fig:U}
\end{figure}

To understand the origin of the JDE in our system, we use the mechanical analogy \cite{BaroneBook, LikharevBook, StrogatzBook}. If we neglect the ac current and thermal fluctuations in Eq.\ \eqref{eq:J1}, then it takes the form of Newton's equation that describes the motion of a particle with mass $\beta$ in the ``washboard'' potential,
\begin{equation} \label{eq:U(phi)}
U(\varphi) = -j_{\mathrm{dc}} \varphi - \cos\varphi - (A/2) \cos(2 \varphi - \tilde{\phi})
\end{equation}
with dissipation; see Fig.~\ref{fig:U}

As shown in Fig.~\ref{fig:U}, the potential shape depends on the current direction (sign of $j_{\mathrm{dc}}$) when $A \sin\tilde{\phi} \neq 0$ (as we will see later this combination is typical for asymmetries). Importantly, the value of $A$ determines the numbers of minima per period: at $A < 1/4$, the potential has only one minimum per period, while at $A > 1/4$, another minimum appears at certain values of $j_{\mathrm{dc}}$.

In the presence of only one minimum per period, we define the oscillation frequencies at the well bottoms $\omega_{A\pm}$, 
curvatures (imaginary frequencies) of the barriers $\omega_{B \pm}$, and barrier heights $\Delta U_{\pm}$.  The `$\pm$' sign in subscripts indicates the branch of the CVC (plus for $j_{\mathrm{dc}} > 0$ and minus for $j_{\mathrm{dc}} < 0$). We measure these frequencies in units of $\omega_{p} $ and the potential barriers in units of $E_{J}$. They can be found in the case of small amplitude of the second harmonic $A \ll 1$:
\begin{gather}
\omega^{2}_{A\pm}= \sqrt{1-j_{\mathrm{dc}}^2}\left(1 \pm A |j_{\mathrm{dc}}|\frac{ 3 - 2j_{\mathrm{dc}}^2}{1 - j_{\mathrm{dc}}^2} \sin \tilde{\phi} \right) 
\notag \\  
+ 2A(1 - j_{\mathrm{dc}}^2)\cos\tilde{\phi}, 
\label{eq:osc} \\
\omega^{2}_{B \pm} = \sqrt{1-j_{\mathrm{dc}}^2}\left(1 \pm A |j_{\mathrm{dc}}|\frac{ 3 - 2j_{\mathrm{dc}}^2}{1 - j_{\mathrm{dc}}^2} \sin\tilde{\phi} \right) 
\notag \\ 
- 2A(1 - j_{\mathrm{dc}}^2)\cos\tilde{\phi}, 
\label{eq:osc at barrier}\\
\Delta U_{\pm} = 2j_{\mathrm{dc}}\arcsin j_{\mathrm{dc}} -\pi |j_{\mathrm{dc}}| 
\notag \\
+ 2 \sqrt{1-j_{\mathrm{dc}}^2}\left(1 \pm  A |j_{\mathrm{dc}}| \sin \tilde{\phi} \right).  \label{eq: Delta U small A}
\end{gather}
Equations \eqref{eq:osc}-\eqref{eq: Delta U small A} are applicable below the critical currents, that is when $|j_\mathrm{dc}| < 1$ and $A \ll 1 -j_{\mathrm{dc}}^2$.

\begin{figure}[t]
 \includegraphics[width=\columnwidth]{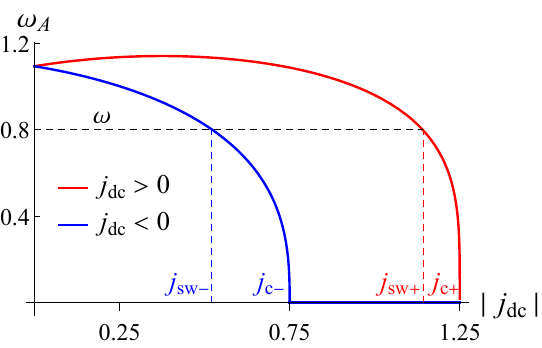}
\caption{Asymmetry of the oscillation frequencies $\omega_{A \pm}$ and the corresponding switching currents $j_{\mathrm{sw} \pm}$ arising as a result of resonant synchronization of internal oscillations with applied external ac irradiation (see Sec.~\ref{subsec: resonance}) at $A = 0.25$ and $\tilde{\phi} = \pi/2$.}
 \label{fig:Omega}
\end{figure}

\begin{figure}[t]
 \includegraphics[width=\columnwidth]{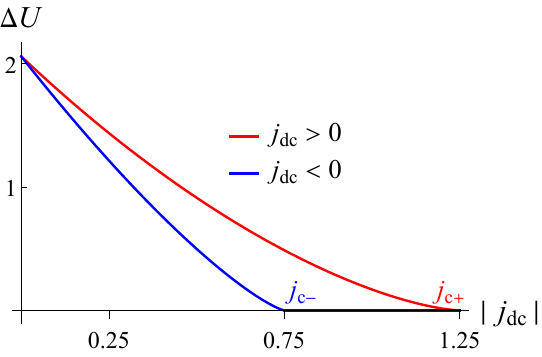}
\caption{Asymmetry of the potential barriers $\Delta U_{\pm}$ at $A = 0.25$ and $\tilde{\phi} = \pi/2$ that leads to exponentially strong asymmetry of the CVC in the low-temperature limit.}
\label{fig:Delta U}
\end{figure}

The same quantities can be found at the ``maximum asymmetry point'' ($\tilde{\phi} = \pi/2$ where $\sin \tilde{\phi} = 1$) without expansion with respect to $A$:
\begin{align}
\omega^2_{A \pm} &= \omega_{B \pm}^2 = \sqrt{4 A (A - j_{\mathrm{dc}})+\sqrt{8 A (A+ j_{\mathrm{dc}})+1}-1} \notag\\
&\times \left( \sqrt{8 A (A + j_{\mathrm{dc}})+1} \right) / 2 \sqrt{2}A \label{eq:osc freq at pi/2},
\\ 
\Delta U_{\pm} 
&= \sqrt{4 A (A-j_{\mathrm{dc}})+\sqrt{8 A (A+j_{\mathrm{dc}})+1}-1} 
\notag \\
&\times \left(\sqrt{8 A (A+ j_{\mathrm{dc}})+1}+3\right)/4 \sqrt{2} A-\pi |j_{\mathrm{dc}}| \notag \\
&+ 2 j_{\mathrm{dc}}  \arcsin \left(\frac{\sqrt{8 A (A+j_{\mathrm{dc}})+1} -1 }{4A}\right) \label{eq: Delta U}. 
\end{align}
In the vicinity of the critical currents ($\Delta j = j_{c \pm} - |j_{\mathrm{dc}}| \ll j_{c\pm}$), they have the following asymptotic behavior: 
\begin{align}
&\Delta U_{\pm} = u_{c \pm}(\Delta j)^{3/2}, \quad u_{c \pm} = \frac{4 \sqrt{2}}{3 \sqrt{1 \pm 4A}}, \label{eq:Asympt U} \\
&\omega_{A \pm} = \omega_{B\pm} = \omega_{c\pm} (\Delta j)^{1/4}, \quad \omega_{c \pm} = 2(1 \pm 4A)^{1/4} \label{eq:Asympt omega}.
\end{align}
Asymmetries of the oscillation frequencies and potential barriers are illustrated in Figs.~\ref{fig:Omega} and~\ref{fig:Delta U}, respectively.

In general, asymmetry of the potential shape leads to different characteristics of the ``particle'' motion [for example, $\overline{v}(j_{\mathrm{dc}})$] for  different motion directions. 

\subsection{Hysteresis of the CVC}

In the case of finite capacitance, $\beta \neq 0$, the CVC may become hysteretic \cite{BaroneBook, LikharevBook, StrogatzBook}. For example, consider the case when $j_{\mathrm{ac}} =0$, $T = 0$, and $A = 0$ (without the JDE). In this case, the CVC is symmetric and depends on the history of current variation. Assume that initially the junction is in the S state, $ 0 < j_{\mathrm{dc}} < j_{c}$. As the current increases, the junction remains in the S state until the current reaches $j_{c}$.  At larger currents, the junction immediately switches to the R state. After that, as the current decreases, the junction returns to the S state only at the retrapping current value $ j_{r} \leq j_{c}$. As a result, in the range $j_{r}< j_{\mathrm{dc}} < j_{c}$ there are two possible branches (S and R), and the system chooses one of them depending on the history.

Similarly, hysteretic behavior occurs in the case of $A \sin \tilde{\phi} \neq 0$ when the JDE is present in the system. The main difference is that in this case the CVC is asymmetric, and the hysteresis is asymmetric too, see Fig.~\ref{fig:hyst}. For example, two different values of the retrapping currents are expected, $j_{r+} \neq j_{r-}$ (for the positive and negative currents, respectively).

\section{Asymmetric CVC}
\label{sec:CVC}

In this section, we investigate manifestations of the JDE in the CVC of the minimal model with nonzero capacitance. As mentioned in the previous section, when $\beta \neq 0$ and  $A \sin\tilde{\phi} \neq 0$, the CVC becomes hysteretic and asymmetric. We consider asymmetries of the characteristic features of the CVC (critical currents $j_{c \pm}$, retrapping currents $j_{r\pm}$, behavior near Ohm's law, etc.) and discuss how capacitance affects them. 

In this section, we assume $T = 0$ and $j_{\mathrm{ac}} =0$ (influence of finite $T$ and $j_{\mathrm{ac}}$ on the CVC is studied in next sections). In the $\beta$ representation, the Josephson equation takes the following form:
\begin{equation} \label{eq: Josephson equation without jac and noise}
\beta \ddot{\varphi} + \dot{\varphi} + \sin \varphi + A \sin(2 \varphi - \tilde{\phi}) =  j_{\mathrm{dc}}.
\end{equation}
Alternatively, in the $\varepsilon$ representation we have
\begin{equation} \label{eq: Josephson equation without jac and noise varepsilon}
\ddot{\varphi} + \varepsilon \dot{\varphi} + \sin\varphi + A \sin(2 \varphi - \tilde{\phi}) =  j_{\mathrm{dc}}.
\end{equation}
In each case, the dot implies derivative with respect to the corresponding time ($\tau$ and $\tilde{\tau}$ in the $\beta$ and $\varepsilon$ representation, respectively).

\subsection{Asymmetry of the critical currents}

Asymmetry of the critical currents  $j_{c+} \neq j_{c-}$ does not depend on the value of the McCumber parameter and is presents even at $\beta = 0$. This asymmetry was studied in detail for different setups in Refs.\ \cite{MikhailovMSThesis2020,Fominov2022.PhysRevB.106.134514,Souto2022.PhysRevLett.129.267702,Gupta2023,Haenel2022arXiv}. Here, we briefly summarize the results for the minimal model. In this model, the diode efficiency
\begin{equation}
 \eta =  \frac{|j_{c+} - j_{c-}|}{j_{c+} + j_{c-}}
\end{equation}
reaches its maximum possible value 
\begin{equation}
    \eta = 1/3,
\end{equation} 
at $A = 0.5$ and $\tilde{\phi} = \pi/2$, with $j_{c+} = 1.5$ and $j_{c-} = 0.75$.

In the limit of small amplitude of the second harmonic $A \ll 1$, the critical currents are given by the following expression:
\begin{equation}
j_{c\pm} = 1 \pm A\sin \tilde{\phi}.
\end{equation}

They can also be found at arbitrary $A$ 
value at $\tilde{\phi} = \pi/2$:
\begin{align} \label{eq: jc}
j_{c+}= 1 + A, \qquad j_{c-} = \begin{cases}
1 - A, \quad  &A < 1/4 \\
A + 1/8A,  &A > 1/4.
\end{cases}    
\end{align}

\subsection{Suppression of the JDE by capacitance}
\label{sec:Suppresion by capacity and harmonic perturbation theory}

\begin{figure}[t]
 \includegraphics[width=\columnwidth]{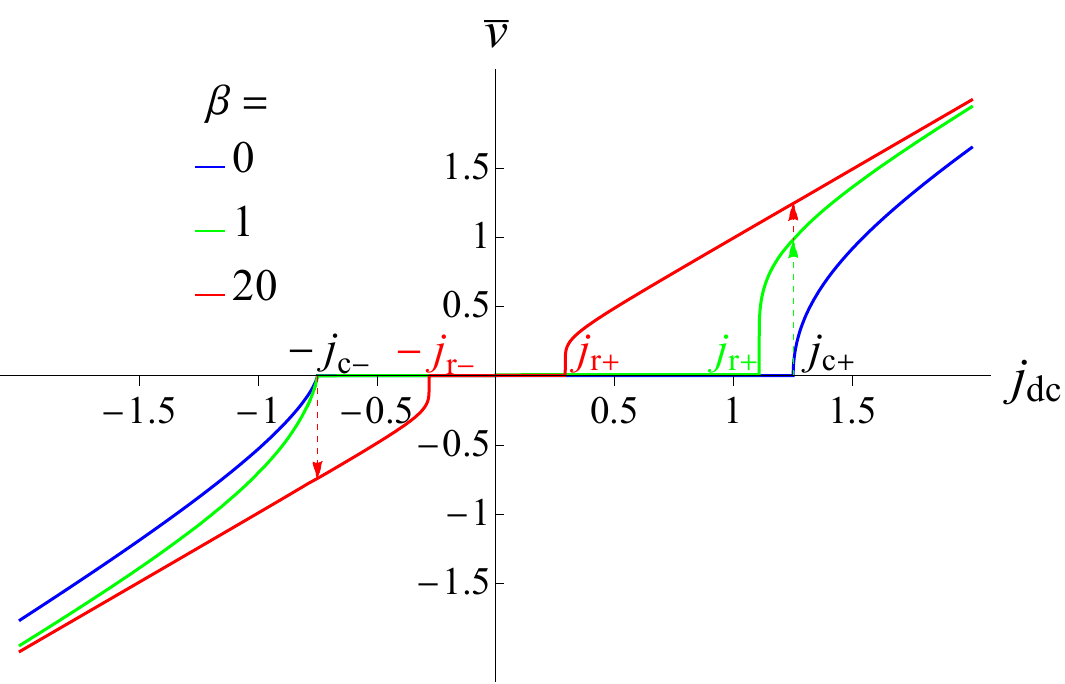}
\caption{Asymmetric CVC at $A = 0.25$ and $\tilde{\phi} = \pi/2$ at different values of the McCumber parameter $\beta$. Arrows indicate switching from the S to R state at the critical currents $j_{c\pm}$. As capacitance increases, the R state of the CVC becomes more symmetric and the JDE weakens.}
 \label{fig:hyst}
\end{figure}

We now investigate how capacitance affects the strength of the JDE. Figure~\ref{fig:hyst} demonstrates the numerically calculated CVC of the minimal model at different $\beta$. As the McCumber parameter $\beta$ increases, asymmetries of the CVC in the R state get suppressed.

In addition to numerical calculations, we investigate the suppression of the JDE by capacitance analytically. We work in the $\beta$ representation and consider the large-capacitance limit near Ohm's law where $\overline{v} = j_{\mathrm{dc}}$ [accurate conditions of applicability will be written below, see Eqs.\ \eqref{eq:large capacitance limit} and \eqref{eq:small capacitance limit}]. We calculate the first nontrivial asymmetric correction to Ohm's law using the ``harmonic perturbation theory'' (HPT) \cite{Fominov2022.PhysRevB.106.134514}. 

Generally, the solution can be represented in the  form
\begin{equation}
 \label{eq:phi as series}
\varphi(\tau) = \overline{v} \tau + \sum \limits_{n = 1}^{\infty} (a_{n} \cos n \overline{v} \tau + b_{n} \sin n \overline{v} \tau), 
\end{equation}
where we take into account that the slope of the linearly growing term is exactly equal to $\overline{v}$, while the remaining part oscillates with frequencies that are multiples of $\overline{v}$ (which is a manifestation of the voltage periodicity with period $2 \pi/ \overline{v}$).

While representation \eqref{eq:phi as series} is general, its parameters $a_n$, $b_n$, and $\overline{v}$ have to be determined from a complicated system of equations, which can be done only numerically.
At the same time, in certain limiting cases, they can be found perturbatively:
\begin{gather}
\varphi(\tau) = j_{\mathrm{dc}} \tau + \sum \limits_{k = 1}^{\infty} \varphi^{(k)}, \quad 
\overline{v}= j_{\mathrm{dc}} + \sum \limits_{k = 1}^{\infty} \overline{v}^{(k)}, \label{eq:series for v} \\
a_{n} =  \sum \limits_{k = 1}^{\infty} a^{(k)}_{n}, \quad b_{n} =  \sum \limits_{k = 1}^{\infty} b^{(k)}_{n}, \notag
\end{gather}
where $\varphi^{(k)}$, $a^{(k)}_{n}$, $b^{(k)}_{n}$, and $\overline{v}^{(k)}$ are contributions to the phase across junction, amplitudes of the $n$th harmonics, and the average voltage in the $k$th order of the perturbation theory. The HPT that we employ below is thus a way to find perturbative expansions \eqref{eq:series for v} for the parameters of the general harmonic representation \eqref{eq:phi as series}.

Substituting Eq.\ \eqref{eq:phi as series} into Eq.\ \eqref{eq: Josephson equation without jac and noise}, expanding the resulting equation into the Fourier series, and solving it, we obtain the CVC, see Appendix~\ref{sec: Appendix A CVC}. Note that we do not expand $\cos n \overline{v} \tau$ and $\sin n \overline{v} \tau$ into series. In this method, corrections to the average voltage $\overline{v}^{(k)}$ appear only due to constant (nonoscillating) terms generated by products of trigonometrical functions. The resulting CVC takes the following form:
\begin{multline}
\label{eq:CVC in large capacity limit}
\overline{v} = j_{\mathrm{dc}} -\frac{4 + A^{2}}{8 \beta^{2} j_{\mathrm{dc}}^3}  + \frac{16 + A^{2} + 24A \beta \cos \tilde{\phi}}{32 \beta^{4} j_{\mathrm{dc}}^5} - \frac{15 {A \sin \tilde{\phi}}}{16 \beta^{4} j_{\mathrm{dc}}^6}. 
\end{multline}
The last term in Eq.\ \eqref{eq:CVC in large capacity limit} is asymmetric [breaks the symmetry $\overline{v}(j_{\mathrm{dc}}) \neq -\overline{v}(-j_{\mathrm{dc}})$] and thus demonstrates the JDE. 
From Eq.\ \eqref{eq:CVC in large capacity limit} we see that the actual parameters of the perturbation theory in the large-capacitance limit  are 
\begin{equation} \label{eq:large capacitance limit}
\beta j_{\mathrm{dc}}^2 \gg 1, \quad \beta j_{\mathrm{dc}} \gg 1,
\end{equation}
and the expansion Eq.\ \eqref{eq:series for v} is carried out according to these parameters. 

The same approach can be used to determine the CVC in the small-capacitance limit:
\begin{equation} \label{eq:small capacitance limit}
j_{\mathrm{dc}}\gg 1, \quad \beta j_{\mathrm{dc}} \ll 1.
\end{equation}
In both the cases of Eqs.\ \eqref{eq:large capacitance limit} and \eqref{eq:small capacitance limit}, we assume that the CVC is approximately given by Ohm's law (this is guaranteed by the first condition in each of the equations). At the same time, the second condition in Eqs.\ \eqref{eq:large capacitance limit} and \eqref{eq:small capacitance limit} determines relative importance of the inertial and dissipative terms in Eq.\ \eqref{eq: Josephson equation without jac and noise}. In the large-capacitance limit, corrections from the inertial term dominates over corrections from the dissipative term, and vice versa in the small-capacitance limit. 

Asymmetric CVC in the small-capacitance limit was calculated in Ref.\ \cite{Fominov2022.PhysRevB.106.134514} at $\beta = 0$. The leading asymmetric term at $\beta \neq 0$ is the same. The result is
\begin{equation} \label{eq:CVC in law capacity limit}
\overline{v} = j_{\mathrm{dc}} - (1 + A^2)/2j_{\mathrm{dc}} - 3 A \sin (\tilde{\phi})/4 j_{\mathrm{dc}}^2.
\end{equation}

Comparing Eqs.\ \eqref{eq:CVC in large capacity limit} and \eqref{eq:CVC in law capacity limit}, we can conclude that asymmetry of the CVC in the large-capacitance limit lies in higher orders of the perturbation theory than in the small-capacitance limit.

\subsection{Asymmetry of the retrapping currents}
\label{sec:Retrapping current}

\begin{figure*}[t]
 \includegraphics[width=0.66\columnwidth]{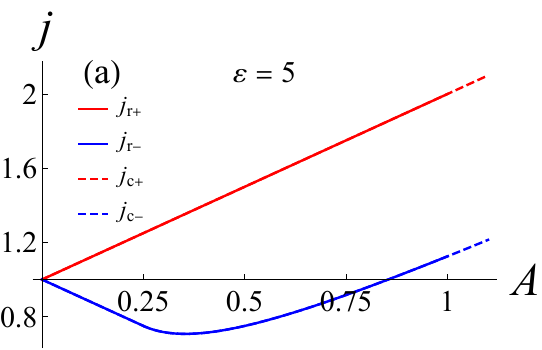}
 \hfill
  \includegraphics[width=0.66\columnwidth]{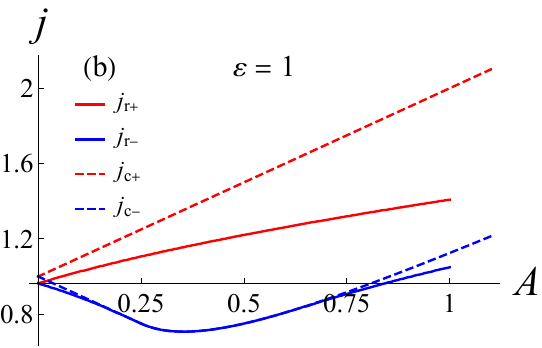}
 \hfill
 \includegraphics[width=0.66\columnwidth]{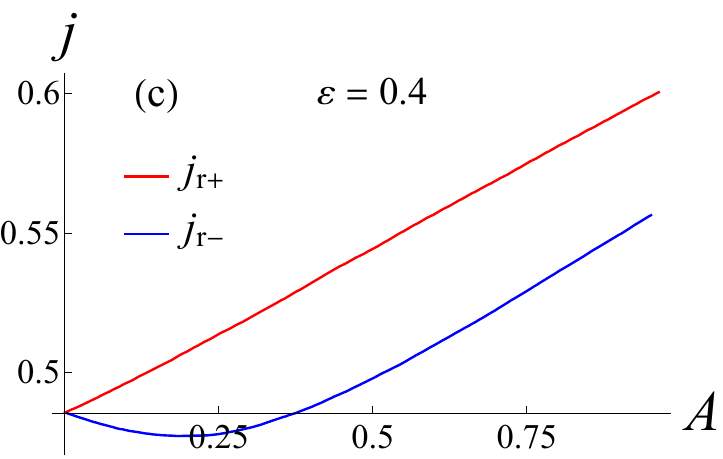}
\caption{Dependence of the retrapping currents $j_{r \pm}$ on the amplitude of the second harmonic $A$ at $\tilde{\phi} = \pi/2$ and at different values of the dissipation factor $\varepsilon = 1/\sqrt{\beta}$. Dashed lines demonstrate the $j_{c\pm}(A)$ dependence. 
(a)~Retrapping currents $j_{r\pm}(A)$ at $\varepsilon = 5$. In this case, for both current directions and within the selected range of $A$ the CVC is nonhysteretic and $j_{r\pm} = j_{c \pm}$. 
(b)~Retrapping currents $j_{r\pm}(A)$ at $\varepsilon = 1$. In this case, for positive current direction the CVC is always hysteretic. At the same time, the CVC for the negative direction can be both hysteretic and nonhysteretic depending on the $A$ value. In the range $A \approx (0.25, 0.75)$, the dashed blue line coincides with the thick blue line, which indicates nonhysteretic CVC. Out of this range, the lines do not coincide, which means that the CVC is hysteretic. 
(c)~Retrapping currents $j_{r\pm}(A)$ at $\varepsilon = 0.4$. In this case, for both current directions the CVC is hysteretic and $j_{r\pm} < j_{c\pm}$. Dashed lines are outside the shown range.}
\label{fig:Retrapping currents}
\end{figure*}

In the previous subsections, we considered the manifestations of the JDE that are present even in the zero-capacitance limit (asymmetry of the critical currents and the CVC near Ohm's law). We now investigate asymmetry of the retrapping currents which have nontrivial values $j_{r} \neq j_{c}$ only at $\beta \neq 0$. Analytical results can be obtained in the limit of weak dissipation, $\varepsilon \ll 1$, and small amplitude of the second harmonic, $A \ll 1$. The $\varepsilon$ representation is convenient in this case. We introduce the energy of the system $E$ and rewrite Eq.\ \eqref{eq:J1 epsilon} as
\begin{gather} \label{eq:E(t)}
E[\varphi(t)] = E_{0} - \varepsilon \int_{\varphi_\mathrm{in}}^{\varphi}\dot{\varphi} d \varphi, \\
  E_{0} = \left. \left( \frac{\dot{\varphi}^2}{2} + U(\varphi) \right)\right|_{\varphi_{\mathrm{in}}},
\end{gather}
where $E_{0}$ and $\varphi_{\mathrm{in}}$ are the initial energy and phase, respectively. The last term in Eq.\ \eqref{eq:E(t)} determines the energy dissipation, and it is parametrically small in this limiting case [$j_{r\pm}$ is also small, see Eq.\ \eqref{eq:j_{r}}].

We therefore employ the perturbation theory with respect to  $\varepsilon \ll 1$ \cite{Chen1988}. The retrapping current corresponds to the separatrix trajectory $\varphi(t)$ that connects two neighboring maxima of the potential ($\varphi_{\max}$ and $\varphi_{\max} \pm 2 \pi$), starts with zero initial velocity and in the final state has the same energy as in the initial one:
\begin{equation} \label{eq:j_{r}}
j_{r\pm} = \frac{\varepsilon}{ 2 \pi} \int_{\varphi_{\max}}^{\varphi_{\max} \pm 2 \pi} \dot{\varphi} d\varphi.
\end{equation}
The perturbation theory implies the following steps. First, we determine the locations of the potential maxima $\varphi_{\max}$ and the corresponding energies $E_{0}$. Second, we express $\dot{\varphi}$ from Eq.\ \eqref{eq:E(t)}, in which we replace $\dot{\varphi}$ in the dissipative term by its value obtained at the previous step. Finally, we substitute the resulting expression for $\dot{\varphi}$ to Eq.\ \eqref{eq:j_{r}} and obtain $j_{r\pm}$.

As a result of this procedure (see Appendix \ref{Appendix:Retrap}), the retrapping currents take the following form:
\begin{equation} \label{eq:ans j_{r}}
   j_{r\pm} = \frac{4 \varepsilon}{\pi}\left(1 - \frac{ A \cos\tilde{\phi} }{3} \pm \frac{ \left(\pi ^2-6\right)  \varepsilon A \sin\tilde{\phi}}{3\pi}\right).
\end{equation}
The first term here coincides with the well-known result \cite{BaroneBook, LikharevBook, Chen1988} for the retrapping current in the large-capacitance limit at $A = 0$.
 
The last term in Eq.\ \eqref{eq:ans j_{r}} is asymmetric. Note that asymmetry $j_{r+} \neq j_{r-}$ is in contrast to Ref.\ \cite{Steiner2023.PhysRevLett.130.177002} where the regime of extremely low dissipation was considered, $\varepsilon \rightarrow 0$. At the same time, asymmetry is proportional to $\varepsilon^2$, hence it arises in the second order of the perturbation theory.

\begin{figure}[t] 
 \includegraphics[width=\columnwidth]{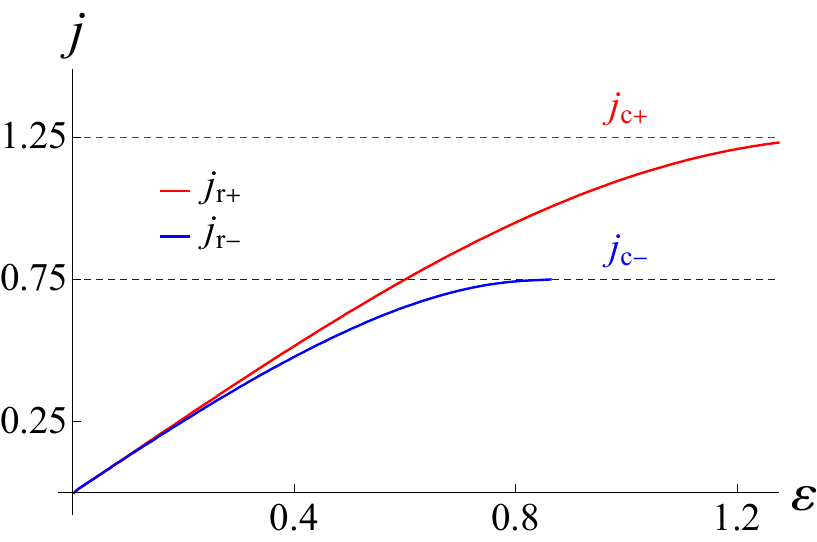}
\caption{Asymmetry of the retrapping currents $j_{r\pm}$ at $\tilde{\phi} = \pi/2$ and $A = 0.25$ for arbitrary values of the dissipation factor $\varepsilon = 1/\sqrt{\beta}$. Dashed lines demonstrate the values of the critical currents $j_{c\pm}$. The CVC for each current direction changes its behavior from hysteretic (at small $\varepsilon$) to nonhysteretic (at large $\varepsilon$) when the thick line crosses the dashed line with the same color. The crossing points are different for different current directions.}
\label{fig:Retrapping current on varepsilon}
\end{figure}

To describe asymmetry of the retrapping currents at arbitrary $\varepsilon$ and $A$, we perform numerical calculations. The results are shown in Figs.~\ref{fig:Retrapping currents} and~\ref{fig:Retrapping current on varepsilon}. 
Asymmetry of the retrapping currents depends both on $A$ and $\varepsilon$, as expected. In the case of strong dissipation, $\varepsilon \gg 1$, the retrapping currents coincide with the critical currents. On the contrary, in the weak-dissipation limit, $\varepsilon \ll 1$, the retrapping currents by themselves are small, $j_{r\pm} \ll j_{c\pm}$; at the same time, asymmetry of their values is weak [see Eq.\ \eqref{eq:ans j_{r}}]. The most interesting case is thus the regime of moderate dissipation, $\varepsilon \sim 1$, in which case the retrapping currents differ from the critical ones (at least, in one current direction) while the asymmetry is still strong.

\subsection{Single-sided hysteresis}

In the RCSJ model with sinusoidal CPR, the CVC is known to become hysteretic (with $j_{r} \neq j_{c}$) only if the dissipation factor is less than a critical value, $\varepsilon \leq \varepsilon_{\mathrm{cr}} \approx 1.18$ \cite{StrogatzBook, BaroneBook, Purrello2020}. In the minimal model, this critical value depends on the current direction, $\varepsilon_{\mathrm{cr}+} \neq \varepsilon_{\mathrm{cr}-}$. This implies that in the range $\max(\varepsilon_{\mathrm{cr}+}, \varepsilon_{\mathrm{cr}-}) > \varepsilon  > \min(\varepsilon_{\mathrm{cr}+}, \varepsilon_{\mathrm{cr}-})$, the hysteresis of the CVC is present in only one current direction, while in the opposite direction the CVC is nonhysteretic \cite{Haenel2022arXiv}. We call this behavior ``single-sided hysteresis''. Manifestations of the single-sided hysteresis are illustrated in Figs.~\ref{fig:hyst} and~\ref{fig:Retrapping currents}. In Fig.~\ref{fig:hyst}, the green curve has a nontrivial value of the retrapping current $j_{r+} \neq j_{c+}$ in the positive current direction and the trivial value in the negative direction, $j_{r-} = j_{c-}$. In Fig.~\ref{fig:Retrapping currents}(b), switching from the hysteretic behavior for both the current directions to the single-sided hysteresis and back is demonstrated as $A$ grows.
 
\begin{figure}[t]
 \includegraphics[width=\columnwidth]{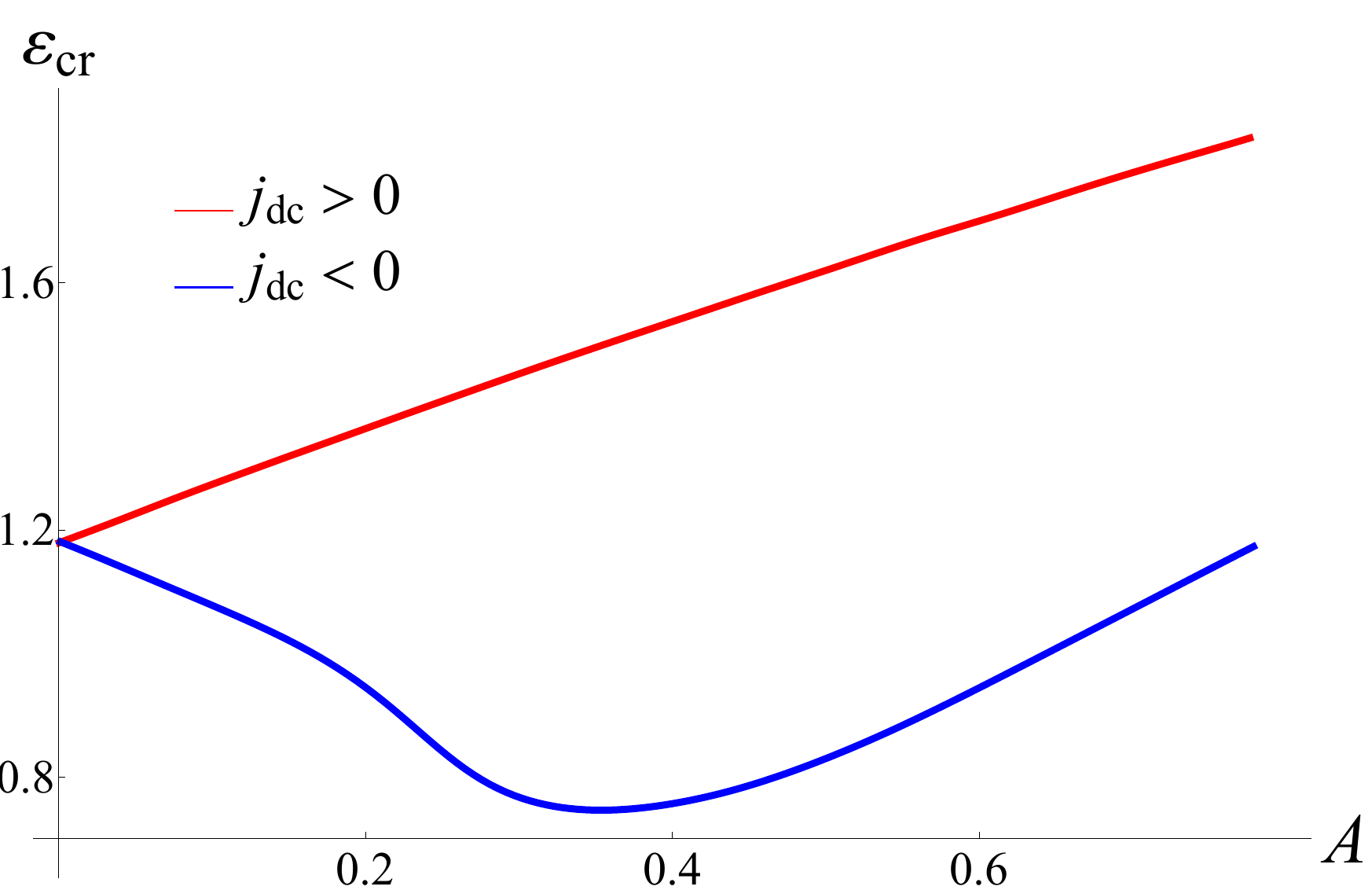}
\caption{Asymmetric dependence of the critical dissipation factor $\varepsilon_{\mathrm{cr}} = 1/\sqrt{\beta_{\mathrm{cr}}}$ that determines the boundaries between the hysteretic ($j_{r} \neq j_{c}$) and nonhysteretic ($j_{r} = j_{c}$) CVCs, on $A$ for different current directions at $\tilde{\phi} = \pi/2$. The region between the red and blue curves is the region of the single-sided hysteresis, where the CVC is hysteretic only in the positive current direction.}
 \label{fig:single sided hyst}
\end{figure}

To investigate the single-sided hysteresis in more detail, we numerically calculate $\varepsilon_{\mathrm{cr}\pm}(A)$, the dependence of the dissipation factor on $A$ for the positive and negative currents. The results shown in Fig.~\ref{fig:single sided hyst} demonstrate that it is possible to observe the single-sided hysteresis in our system in a wide range of $\varepsilon$ (or $\beta$).
 
\section{JDE in the presence of microwave irradiation}
\label{sec:irradiation}

In this section, we discuss the CVC of the asymmetric SQUID under the influence of external microwave irradiation that generates the ac current $j_{\mathrm{ac}} \neq 0$. Except in Sec.~\ref{subsec: resonance}, we consider the zero-temperature limit, corresponding to $\xi = 0$ in the Josephson equation in the $\beta$ representation:
 \begin{equation} \label{eq: J1 with ac}
\beta \ddot{\varphi} + \dot{\varphi} + \sin\varphi + A \sin(2 \varphi - \tilde{\phi}) =  j_{\mathrm{dc}} + j_{\mathrm{ac}} \cos(\omega \tau + \delta).
\end{equation}

\subsection{Asymmetry of the critical currents at 
\texorpdfstring{$j_{\mathrm{ac}} \neq 0$}{j\_ac != 0}}

First, we investigate the effect of external ac irradiation on asymmetry of the critical currents.

\subsubsection{Quasistationary regime}

At $\omega \ll \omega_{A \pm}(j_{\mathrm{dc}})/\sqrt{\beta}$ (the period of oscillations in the S state is much smaller than the period of the ac signal) and $\omega \ll 1$ (the relaxation time is much smaller than the period of the ac signal), the ac current changes very slowly compared to the junction dynamics. The phase dynamics is then quasistationary and the ac and dc currents simply add up \cite{BaroneBook, Krasnov2024.PhysRevApplied.22.024015, Souto2024.PhysRevResearch.6.L022002}. In this case, the critical currents $j_{\mathrm{mw}c\pm}$ under external irradiation are given by the simple expression
 \begin{equation}
 j_{\mathrm{mw}c \pm} = j_{c\pm} - j_{\mathrm{ac}}.
 \end{equation}
As a result, the diode efficiency in the presence of microwave irradiation is given by
 \begin{equation}
 \eta_{\mathrm{mw}} = \frac{|j_{\mathrm{mw}c +} - j_{\mathrm{mw}c -}|}{j_{\mathrm{mw}c +} + j_{\mathrm{mw}c -}} = \frac{|j_{c+}- j_{c-}|}{j_{c+} + j_{c-} - 2 j_{\mathrm{ac}}}.
 \end{equation}
The diode efficiency can thus be enhanced by external microwave irradiation \cite{Souto2024.PhysRevResearch.6.L022002}. It reaches its maximum possible value  $\eta_{\mathrm{mw}} = 1$ at $j_{\mathrm{ac}} = \min(j_{c+}, j_{c -})$.

\subsubsection{Zero-voltage step}

In the case of nonquasistationary dynamics, to obtain analytical results, we assume that $J(\varphi) \ll j_{\mathrm{ac}}$ and employ the perturbation theory with respect to this smallness:
\begin{equation}
    \varphi(\tau) = \varphi^{(0)}(\tau) + \varphi^{(1)}(\tau) + \dots 
\end{equation}
We consider the S state with $\overline{v} = 0$, which means that $\varphi$ does not grow in time.

In the zeroth order of the perturbation theory, the supercurrent and the corresponding dc current in the S state are small. The phase dynamics is then determined only by the external ac current:
\begin{gather}
\beta \ddot{\varphi}^{(0)} + \dot{\varphi}^{(0)} = j_{\mathrm{ac}} \cos(\omega \tau + \delta), \notag \\
\varphi^{(0)}(\tau) = \phi_0 + a\sin(\omega \tau + \tilde{\delta}), 
\label{eq: ac phi 0}\\
a = \frac{j_{\mathrm{ac}}}{\omega \sqrt{(\beta \omega)^2 + 1}}, \quad \tilde{\delta} = \delta + \arctan \beta \omega \notag,
\end{gather}
where $\phi_0$ is the (arbitrary) initial phase across the junction.

Substituting $\varphi^{(0)}(\tau)$ from Eq.\ \eqref{eq: ac phi 0} to the supercurrent term in Eq.\ \eqref{eq: J1 with ac}, in the first order of the perturbation theory we obtain
 \begin{gather}
\beta \ddot{\varphi}^{(1)} + \dot{\varphi}^{(1)} + J(\tau) = j_{\mathrm{dc}}, \label{eq: phi 2 order} \\
J(\tau) = \sum_{n = -\infty}^{\infty} (-1)^{n} [J_{n}(a) \sin(\phi_0 - n\omega \tau - n \tilde{\delta})  \notag \\
 + A J_{n}(2a)\sin(2\phi_0 - n\omega \tau - n \tilde{\delta} - \tilde{\phi})],
 \end{gather}
where $J_{n}$ are the Bessel functions of the first kind. 

Oscillating terms in Eq.\ \eqref{eq: phi 2 order} are responsible for oscillating corrections to the phase, while the nonoscillating term [time-averaged current $\overline{J}(\tau)$] corresponds to the dc current in the S state, $j_{\mathrm{dc}} =\overline{J}(\tau)$. This value depends on $\phi_0$, and the critical currents in this case are thus given by
\begin{equation} \label{eq: crit currents with ac}
j_{\mathrm{mw}c\pm} =\bigl| \underset{\phi_0}{\maxmin} [J_{0}(a) \sin\phi_0 + A J_{0}(2a) \sin(2 \phi_0 - \tilde{\phi})]\bigl|. 
 \end{equation}
As we see, Eq.\ \eqref{eq: crit currents with ac} reproduces the result for the critical currents in the minimal model but with renormalized amplitudes of the first and second Josephson harmonics. Equation \eqref{eq: crit currents with ac} is applicable if $J_{n}(a)$ and $J_{n}(a)/\omega$ are small, for example at $j_{\mathrm{ac}} \gg 1$ and $\omega \sim 1$.

Despite decreased absolute values of the critical currents, in this case the diode efficiency can still reach its maximum possible value for the minimal model $\eta_\mathrm{mw} = 1/3$ at $\tilde{\phi} = \pi/2$ if $A J_{0}(2a) = J_{0}(a)/2$.

\subsection{Asymmetry of the Shapiro steps}

It is well known fact that the external current can synchronize with the internal oscillations of the phase. As a result, the peculiarities called the Shapiro steps arise in the R state in the CVC at $\overline{v} = (n/k) \omega$, where $n$ and $k$ are integers \cite{BaroneBook, Shapiro1963.PhysRevLett.11.80}. Below, we analytically investigate  asymmetry of the height of the first Shapiro steps $j_{\pm 1}$ (corresponding to $\overline{v} = \pm \omega$) in the two limiting cases: 
(i)~large-capacitance limit [defined by Eq.\ \eqref{eq:large capacitance limit}] at $j_{\mathrm{ac}} \ll 1$, and 
(ii)~small-capacitance limit [defined by Eq.\ \eqref{eq:small capacitance limit}]. 
 
We employ a slight modification of the HPT described in Sec.~\ref{sec:Suppresion by capacity and harmonic perturbation theory}. The difference is that we now find the dependence $j_{\mathrm{dc}}(\overline{v})$ instead of $\overline{v}(j_{\mathrm{dc}})$. We use the expansion \eqref{eq:phi as series} with 
\begin{gather} \label{eq:phi as series ac}
\varphi(\tau) = \overline{v}\tau + \sum \limits_{k = 1}^{\infty} \varphi^{(k)}, \quad 
j_\mathrm{dc}= \overline{v} + \sum \limits_{k = 1}^{\infty} j_{\mathrm{dc}}^{(k)}. 
\end{gather}
Employing the modified HPT (see Appendix~\ref{sec: Appendix A Shapiro}), we obtain asymmetries of the heights of the first Shapiro steps in the small-capacitance limit:
\begin{gather}
  j_{\pm 1} = j_{\mathrm{sym}} \pm j_{\mathrm{asym}},
\notag \\ 
  j_{\mathrm{sym}} = j_{\mathrm{ac}}/ \omega , \quad j_{\mathrm{asym}}  = 9 j_{\mathrm{ac}} A \sin(\tilde{\phi}) / 4 \omega^2.
 \label{eq:Shapiro steps in small capacitance limit}
\end{gather}
Similarly, we find the same quantities in the large-capacitance limit:
\begin{equation} \label{eq:Shapiro steps in the large capaticance limit}
     j_{\mathrm{sym}} = j_{\mathrm{ac}}/ \beta \omega^2 , \quad j_{\mathrm{asym}}  = 45 j_{\mathrm{ac}} A \sin(\tilde{\phi}) /16 \beta^3 \omega^5.
\end{equation}

In both the results \eqref{eq:Shapiro steps in small capacitance limit} and \eqref{eq:Shapiro steps in the large capaticance limit}, we keep only the leading terms in the symmetric, $j_{\mathrm{sym}}$, and asymmetric, $j_{\mathrm{asym}}$, parts of the step heights. Similarly to  Sec.~\ref{sec:Suppresion by capacity and harmonic perturbation theory}, asymmetry in the large-capacitance limit arises in a higher order of the HPT than in the small-capacitance limit.

\begin{figure}[t]
 \includegraphics[width=\columnwidth]{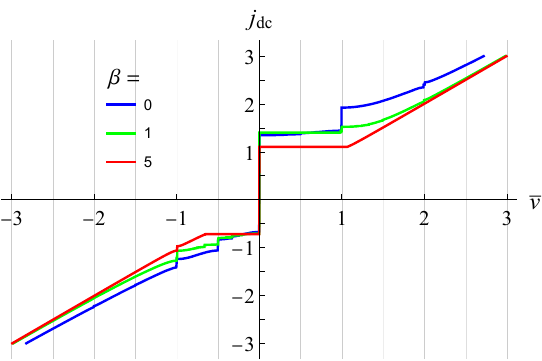}
\caption{Asymmetric Shapiro steps in the R state at $A = 0.5$, $j_{\mathrm{ac}} = 0.5$, $\tilde{\phi} = \pi/2$, and different $\beta$. Asymmetry of the main integer ($\overline{v} = \pm1$) and half-integer ($\overline{v} = \pm 1/2$) Shapiro steps are most clearly visible. As $\beta$ increases, this asymmetry weakens. This is a manifestation of the suppression of the JDE in the R state by capacitance.}
 \label{fig:Shapiro steps}
\end{figure}

In addition to the analytical results for the first steps, we perform numerical calculations to study asymmetry of all possible steps. The results are shown in Fig.~\ref{fig:Shapiro steps}. Both the heights of the Shapiro steps and asymmetry of their heights decrease as $\beta$ increases. This is a manifestation of the suppression of the JDE in the R state by capacitance.

\subsection{Asymmetry of the resonance frequencies}
\label{subsec: resonance}

As discussed in Sec.~\ref{sec:U}, the general form of the potential and, in particular, the oscillation frequency at the well bottom $\omega_{A\pm}$, depend on the current direction. When the system is exposed to external microwave radiation, in addition to the appearance of the Shapiro steps in the R state, resonances in the S state might occur at $\tilde{\omega} \simeq \omega_{A \pm}(j_{\mathrm{dc}})$. We confine our attention to the regime of weak perturbation of the Brownian motion of the particle in the potential well at $T \neq 0$ by weak ac signal \cite{Larkin1986, Devoret1987.PhysRevB.36.58, Krasnov2024.PhysRevApplied.22.024015}. In this regime, the Josephson resonance manifests itself in the great enhancement (with factors $\gamma_{\pm}$) of the thermal escape rates from the S to R state under the influence of weak ac irradiation in junctions with high quality factors $Q_{\pm}(j_{\mathrm{dc}}) = \omega_{A\pm}(j_{\mathrm{dc}}) \omega_{p} RC \gg 1$:
\begin{equation} \label{eq:Gamma}
\Gamma_{\mathrm{mw} \pm} (j_{\mathrm{dc}},\tilde{\omega}) = \gamma_{\pm} (j_{\mathrm{dc}}, \tilde{\omega}) \Gamma_{\pm}(j_{\mathrm{dc}}),
\end{equation}
where $\Gamma_{\pm}$ are the thermal escape rates in the absence of ac irradiation (see Sec.~\ref{sec:thermal escapes}) and $\Gamma_{\mathrm{mw} \pm}$ are the modified escape rates in its presence. Due to difference in the oscillation frequencies $\omega_{A \pm}(j_{\mathrm{dc}})$, the enhancement factors for different current directions would differ: $\gamma_{+}(j_{\mathrm{dc}}, \tilde{\omega}) \neq \gamma_{-}(j_{\mathrm{dc}}, \tilde{\omega})$. 

Moreover, it is known that the  junction can actually switch from the S to R state at currents below the critical values due to thermal escapes of the particle from the potential well. It is possible to find the distribution of the corresponding switching currents. In the absence of external ac irradiation, this distribution has peaks at the thermal switching current values. In the presence of microwave irradiation, new peaks appear in the switching current distribution due to the Josephson plasma resonances in the S state \cite{Gronbech-Jensen2004.PhysRevLett.93.107002, Krasnov2024.PhysRevApplied.22.024015}. Positions of these peaks can be found from the resonance condition $\tilde{\omega} = \omega_{A \pm}(j_{\mathrm{sw}\pm})$. In the case of small second harmonic, from Eq.\ \eqref{eq:osc} we obtain
\begin{equation}
j_{\mathrm{sw}\pm} = \sqrt{1 - \tilde{\omega}^4} + \frac{2 A \tilde{\omega}^{6} \cos\tilde{\phi}}{\sqrt{1 - \tilde{\omega}^{4}}} \pm A (1 + 2 \tilde{\omega}^4)\sin\tilde{\phi},
\end{equation}
which is applicable not too close to the critical currents, i.e., when Eq.\ \eqref{eq:osc} is applicable and $A \ll 1 - \tilde{\omega}^4$.

Asymmetry of the switching currents is illustrated in Fig.~\ref{fig:Omega}. The values of the switching currents can significantly differ for different current directions in the presence of ac irradiation due to the difference in the oscillation frequencies at the well bottom.

\section{Thermal fluctuations}
\label{sec:Tfluct}

In previous sections, we mainly discussed the CVC in the absence of thermal fluctuations (i.e., at $T = 0$). In this section, we take them into account in Eq.\ \eqref{eq:J1} in the simplified case of $j_{\mathrm{ac}} = 0$:
 \begin{equation} \label{eq:thermalJ1}
\beta \ddot{\varphi} + \dot{\varphi} + \sin\varphi + A \sin(2 \varphi - \tilde{\phi}) =  j_{\mathrm{dc}} + \xi(\tau).
 \end{equation}
The presence of thermal fluctuations causes escapes of the particle from the potential minima due to thermal activation. This results in nonzero escape rates from the S state and leads to modification of the CVC \cite{BaroneBook, LikharevBook, Ambegaokar1969.PhysRevLett.22.1364, Ivanchenko1968}. We consider to the low-temperature limit in the sense that $\theta \ll \Delta U_{\pm}$. In this case, the escape time is much larger than the sliding time after thermal activation.

\subsection{CVC at \texorpdfstring{$T \neq 0$}{T != 0} in the zero-capacitance limit}

We start our consideration from the zero-capacitance limit $\beta = 0$. In this case, it is possible to obtain analytical expression for asymmetric CVC in the asymmetric potential of general form
$U_{\mathrm{asym}}(\varphi) = -j_{\mathrm{dc}} \varphi - \sum \limits_{k = 1}^{\infty} A_{k} \cos(k \varphi - \tilde{\phi}_{k})$. For convenience, below in this section we assume that $j_{\mathrm{dc}} > 0$, and then to describe the negative branch of the CVC we use the symmetry
\begin{equation} \label{eq:symmetry}
v(-j_{\mathrm{dc}}, \tilde{\phi}_{k}) = - v(j_{\mathrm{dc}}, -\tilde{\phi}_{k}).
\end{equation}
Technically, this method implies considering two potentials with $j_{\mathrm{dc}} > 0$ (for the positive and negative branch of the CVC, respectively): 
\begin{equation} \label{eq:Uasym}
U_{\mathrm{asym\pm}}(\varphi) = -j_{\mathrm{dc}} \varphi - \sum \limits_{k = 1}^{\infty} A_{k} \cos(k \varphi \mp \tilde{\phi}_{k}), \quad j_\mathrm{dc} > 0.
\end{equation}

Below, we apply the method by Ambegaokar and Halperin \cite{Ambegaokar1969.PhysRevLett.22.1364}. We follow the standard procedure and convert Eq.\ \eqref{eq:thermalJ1} to the Fokker-Planck equation for the stationary distribution function $\sigma_{\mathrm{st}}(\varphi)$:
\begin{equation}
\partial \sigma_{\mathrm{st}}/\partial \varphi + (1/{\cal{\theta}})(\partial U_{\mathrm{asym\pm}}/\partial \varphi)\sigma_{\mathrm{st}} = Q_{\pm}, 
\end{equation}
where $Q_{\pm}(j_{\mathrm{dc}})$ is a constant, which should be found from the normalization condition and periodicity of the distribution function:
\begin{equation}
    \int^{2 \pi}_{0}\sigma_{\mathrm{st}} (\varphi) d\varphi  = 1, \quad \sigma_{\mathrm{st}}(\varphi + 2 \pi) = \sigma_{\mathrm{st}}(\varphi).
\end{equation}

In this language, the expressions for the average voltages in the positive and negative branches are given by
\begin{equation} \label{eq:fluctV}
\langle \overline{v}_{\pm}(j_{\mathrm{dc}}) \rangle = \pm 2 \pi \theta Q_{\pm}(j_{\mathrm{dc}}).
\end{equation}

The expression for $Q_{\pm}$ can be obtained by taking the integral
\begin{equation} \label{eq:Z}
 \frac{1}{Q_{\pm}}= \! - \! \int 
\limits^{2 \pi}_{0} d\varphi \exp\left( -\frac{U_{\mathrm{asym\pm}}(\varphi)}{\theta}\right) \int \limits^{\infty}_{\varphi}dx \exp\left(\frac{U_{\mathrm{asym\pm}}(x)}{\theta}\right).
\end{equation}
To calculate the integrals in Eq.\ \eqref{eq:Z}, we employ the saddle-point approximation. In the low-temperature limit, the maxima and minima of the potential are well separated from each other, and both the integrals are determined by small vicinities of the potential extrema (minima for the external integral and maxima for the internal one). We denote the locations of those maxima and minima as $\varphi_{\max\pm,m}$ and $\varphi_{\min\pm,n}$, respectively. The result of integration can then be written as
\begin{multline} \label{eq:Zcalcul}
\frac{1}{Q_{\pm}} = {\sum \limits_{n,m}}' \sqrt{\frac{2 \pi \theta}{|U_{\mathrm{asym \pm}}''(\varphi_{\min\pm,n})|}} \sqrt{\frac{2 \pi \theta}{|U_{\mathrm{asym\pm}}''(\varphi_{\max\pm,m})|}} 
\\
\times \exp\left[-U_{\mathrm{asym\pm}}(\varphi_{\min\pm,n}) + U_{\mathrm{asym\pm}}(\varphi_{\max\pm,m})\right],
\end{multline}
where the prime sign indicates that at fixed $n$ the sum is taken over such $m$ that satisfy the following relation: $\varphi_{\max\pm,m} > \varphi_{\min\pm,n}$ (so that the corresponding maxima are within the integration region of the internal integral). Equations \eqref{eq:fluctV} and \eqref{eq:Zcalcul} determine the asymmetric CVC due to thermal fluctuations for arbitrary asymmetric potential in the low-temperature limit at currents below the critical one.

In the minimal model, in the presence of only one minimum per period,\footnote{In the presence of two nontrivial minima per period, the stochastic dynamics become more complicated due to interplay between different S states of the junction \cite{Zonda2024.PhysRevB.110.054306}.} one can rewrite the general expression \eqref{eq:Zcalcul} in notations of  Sec.~\ref{sec:U} and obtain the asymmetric CVC in the following form:
\begin{equation} \label{eq:answer Ambegokar pi/2}
\langle \overline{v}_{\pm}(j_{\mathrm{dc}}) \rangle = \pm 2 \omega_{A\pm} \omega_{B\pm} \sinh\!\left(\frac{j_{\mathrm{dc} } \pi}{\theta}\right)
\exp\!\left(\frac{-\Delta U_{\pm} -  j_{\mathrm{dc}} \pi}{\theta} \right) \! .  
\end{equation}
At $\tilde{\phi} = \pi/2$, the quantities entering Eq.\ \eqref{eq:answer Ambegokar pi/2} are given by Eqs.\ \eqref{eq:osc freq at pi/2} and \eqref{eq: Delta U}. At the same time, the most interesting case $\theta \ll A \ll 1$, in which, despite the smallness of the second harmonic, asymmetry is exponentially strong, can be considered at arbitrary $\tilde{\phi}$. In this limit, we keep only the leading term of the first order with respect to $A$ and obtain (for more details, see Appendix~\ref{sec:Ambegaokar for min model})
\begin{multline} \label{eq:answer Ambegaokar}
 \langle \overline{v}_{\pm}(j_{\mathrm{dc}}) \rangle = \pm 2 \sinh \left(\frac{j_{\mathrm{dc}} \pi}{\theta}\right)\sqrt{1-j_{\mathrm{dc}}^2} \\
\times \left(1 \pm A j_{\mathrm{dc}}\frac{ 3 - 2j_{\mathrm{dc}}^2}{1 - j_{\mathrm{dc}}^2} \sin\tilde{\phi}\right) \\ \times \exp \left\{-\frac{2}{\theta} \left[\sqrt{1-j_{\mathrm{dc}}^2}\left(1  \pm  A j_{\mathrm{dc}} \sin\tilde{\phi} \right)+j_{\mathrm{dc}} \arcsin j_{\mathrm{dc}} \right]\right\} \! .
\end{multline}

Note that in Eqs.\ \eqref{eq:answer Ambegokar pi/2} and \eqref{eq:answer Ambegaokar}, asymmetric terms appear both in the prefactor (due to asymmetry of the oscillation frequencies $\omega_{A \pm}$ and $\omega_{B \pm}$) and in the exponent (due to asymmetry of the potential barrier heights $\Delta U_{\pm}$). As expected, asymmetry of the potential barriers leads to exponentially strong asymmetry of the CVC.

\subsection{Escape rates at nonzero capacitance}
\label{sec:thermal escapes}

At $\beta \neq 0$, there is no simple expression for the CVC in the presence of thermal fluctuations. However, it is possible to obtain analytical results for the escape rates from a potential well of the arbitrary asymmetric potential $U_{\mathrm{asym} \pm}$ defined by Eq.\ \eqref{eq:Uasym}, under the assumption that it has only one maximum $\varphi_{\max\pm}$ and minimum $\varphi_{\min\pm}$ per period. As in the previous subsection, we employ the Fokker-Planck equation for the distribution function $\sigma(\varphi, v, \tilde{\tau})$, which in this case is nonstationary:
\begin{equation} \label{eq:FP}
\frac{\partial \sigma}{\partial \tilde{\tau}}  = - \frac{\partial}{\partial \varphi }\left(p \sigma\right) + \frac{\partial}{\partial p}\left[\left(\frac{\partial U_{\mathrm{asym} \pm}}{\partial \varphi} + \varepsilon p \right)\sigma \right] + \varepsilon \theta \frac{\partial^2 \sigma}{\partial p^2},
\end{equation}
where we use the $\varepsilon$ representation for the ease of comparison with previous works \cite{Steiner2023.PhysRevLett.130.177002, Buttiker1983.PhysRevB.28.1268}.

Thermal fluctuations stimulate escapes of the particle from the potential minima. The states in the potential wells thus become metastable with finite lifetimes $\tau_{l\pm}$. This lifetime (in units of $\omega_{p}^{-1}$) was found in Refs.\ \cite{Kramers1940,Mel'nikov1991, Buttiker1983.PhysRevB.28.1268} in a broad range of the McCumber parameters, from the overdamped to underdamped regimes. Generalizing this results to the case of the asymmetric potential, we obtain
\begin{equation} \label{eq:general lifetime}
\tilde{\tau}_{l\pm}^{-1} = \frac{\omega_{A\pm}}{2 \pi \omega_{B\pm}}\left[\left(\frac{\varepsilon^2}{4} + \omega_{B\pm}^2\right)^{1/2} - \frac{\varepsilon}{2}\right] \exp{\left(-\frac{\Delta U_{\pm}}{\theta}\right)},
\end{equation}
where $\omega_{A\pm}$ is the oscillation frequency at the well bottom, $\omega_{B\pm}$ is the imaginary oscillation frequency of the barrier, and $\Delta U_{\pm}$ is the height of the potential barrier of $U_{\mathrm{asym} \pm}$. 
In the minimal model, these quantities are given by Eqs.\ 
\eqref{eq:osc}-\eqref{eq: Delta U}. The general expression \eqref{eq:general lifetime} can be simplified in two limiting cases:
\begin{equation} \label{eq: lifetime }
\tilde{\tau}_{l\pm}^{-1} = 
\begin{cases}
\frac{\omega_{A\pm} \omega_{B \pm}}{2 \pi \varepsilon} \exp{\left(-\frac{\Delta U_{\pm}}{\theta}\right)}, &\quad \varepsilon \gg 1, 
\\
 \frac{\omega_{A\pm}}{2 \pi} \exp{\left(-\frac{\Delta U_{\pm}}{\theta}\right)}, 
 &\quad \varepsilon \ll 1.
\end{cases}
\end{equation}

The above results \eqref{eq:general lifetime} and \eqref{eq: lifetime } are applicable if the dissipation is not extremely small: $\varepsilon \gg  \omega_{B\pm} \theta/ \Delta U_{\pm}$ \cite{Mel'nikov1991}. When this condition is violated, it is necessary to take into account the depopulation below the barrier top. In the very-large-capacitance limit (so-called extremely underdamped regime), the switching rate was found in Refs.\ \cite{Kramers1940,Steiner2023.PhysRevLett.130.177002}:
\begin{equation} \label{eq:lifetime very large capacitance limit}
\tilde{\tau}_{l\pm}^{-1} = \frac{\varepsilon  \omega_{A\pm} S_{\pm}}{2 \pi \theta} \exp{\left(-\frac{\Delta U_{\pm}}{\theta}\right)}, \quad \varepsilon \ll \theta/S_{\pm}.
\end{equation}
Here, $S_{\pm} = S_{\pm}(j_{\mathrm{dc}})$ is the action of the separatrix motion corresponding to the trajectory that starts at a maximum $\varphi_{\max\pm}$ with zero initial velocity and after one oscillation in the potential well [with turning point $\varphi_{\mathrm{tp}\pm}$ such that $U_{\mathrm{asym}\pm}(\varphi_{\mathrm{tp}\pm}) = U_{\mathrm{asym}\pm}(\varphi_{\max\pm})$] returns back to the maximum:
\begin{equation}
S_{\pm} = 2 \int \limits_{\varphi_{\mathrm{tp}\pm}}^{\varphi_{\max\pm}} \sqrt{2\left[U_{\mathrm{asym\pm}}\left(\varphi_{\max\pm}\right) - U_{\mathrm{asym\pm}}\left(\varphi\right)\right]} d\varphi.
\end{equation}

Note that in all the cases above, asymmetric lifetime can be written in the following form:
\begin{equation} \label{eq:general form of lifetime}
\tilde{\tau}^{-1}_{l\pm} = \frac{\tilde{\omega}_{\mathrm{att}\pm}}{2 \pi} \exp\left(- \frac{\Delta U_{\pm}}{\theta} \right),
\end{equation}
where $\tilde{\omega}_{\mathrm{att}\pm}$ is the effective attempt frequency of the thermal activation process. 

To emphasize the physical meaning of $\tilde{\tau}_{l\pm}^{-1}$, we note that it is nothing but the escape rates $\Gamma_{\pm}$ entering Eq.\ \eqref{eq:Gamma}. These rates are also related to the average voltage by the simple expression in the overdamped regime:
\begin{equation} \label{eq:voltage and escape rates}
\langle\overline{v}_{\pm}(j_{\mathrm{dc}})\rangle = \pm 2 \pi \varepsilon (\Gamma_{\pm} - \Gamma^{\leftarrow}_{\pm}), \quad \Gamma^{\leftarrow}_{\pm} = \exp \left(-\frac{2 j_{\mathrm{dc}} \pi}{\theta} \right)\Gamma_{\pm}.
\end{equation}
Here, $\Gamma_{\pm}$ is the escape rate from the potential well to the right while $\Gamma^{\leftarrow}_{\pm}$ is the escape rate to the left.
Equation \eqref{eq:voltage and escape rates} demonstrates the equivalence between Eqs.\ \eqref{eq:answer Ambegokar pi/2} and \eqref{eq: lifetime }. Note that Eqs.\ \eqref{eq:answer Ambegokar pi/2} and \eqref{eq: lifetime } are written in different representations and this is why the $\varepsilon$ factor appears in Eq.\ \eqref{eq:voltage and escape rates}.

\subsection{Asymmetry of the switching currents}

Escapes from the potential wells due to thermal activation processes lead to switching from the S to R state of the junction at switching currents $j_{\mathrm{sw}\pm} < j_{c\pm}$. Assume that the current is initially zero, and then it slowly increases linearly with time \cite{LikharevBook}:
\begin{equation}
j_{\mathrm{dc}}(\tilde{\tau}) = a \tilde{\tau}, \quad a \tilde{\tau}_{l \pm} \ll 1.  
\end{equation}
The probabilities to remain in the potential well when the current reaches the value $j_{\mathrm{dc}}$ are then given by
\begin{equation}
 \sigma_{\mathrm{sw}\pm}(j_{\mathrm{dc}}) =  \exp\left[-\frac{1}{a} \int_{0}^{j_{\mathrm{dc}}} \tilde{\tau}_{l\pm}^{-1}\left(j\right) d j \right].
\end{equation}
In the low-temperature limit, due to the fast decrease of the exponential in $\tilde{\tau}_{l\pm}$, the probability at this value takes the form
\begin{equation}
\sigma_{\mathrm{sw} \pm}(j_{\mathrm{sw\pm}}) =  \exp\left(-\frac{\theta}{a} \tilde{\tau}_{l\pm}^{-1}(j_{\mathrm{sw\pm}}) \left|\frac{d \Delta U_{\pm}(j_{\mathrm{dc}})}{d j_{\mathrm{dc}}}\right|^{-1}_{j_{\mathrm{dc}} = j_{\mathrm{sw\pm}}}\right).
\end{equation}
Following Ref.\ \cite{Steiner2023.PhysRevLett.130.177002}, we define the switching currents from the relation
\begin{equation}
\sigma_{\mathrm{sw \pm}}(j_{\mathrm{sw \pm}}) = 1/2.
\end{equation}
The implicit expression for the switching currents then takes the form
\begin{equation} \label{eq:sw current}
a \ln 2 = \theta \tilde{\tau}^{-1}_{l \pm} (j_{\mathrm{sw} \pm}) \left| \frac{d \Delta U_{\pm}(j_{\mathrm{dc}})}{d j_{\mathrm{dc}}}\right|^{-1}_{j_{\mathrm{dc}} = j_{\mathrm{sw\pm}}}.
\end{equation}
In the most general case, the lifetime is given by Eq.\ \eqref{eq:general form of lifetime} and the equation for the switching current takes the form
\begin{equation} \label{eq:implitic exprerssion for sw currents}
\frac{\Delta U_{\pm}(j_{\mathrm{sw}\pm})}{\theta} = \ln\left(\frac{\theta \tilde{\omega}_{\mathrm{att}\pm}(j_{\mathrm{sw}})}{2 \pi a \ln 2} \left|\frac{d \Delta U_{\pm} (j_{\mathrm{dc}})}{d j_{\mathrm{dc}}}\right|^{-1}_{j_{\mathrm{dc}} = j_{\mathrm{sw}\pm}}\right).
\end{equation}
To obtain explicit expressions for the switching currents, one should substitute here asymptotic expressions for $\Delta U_{\pm}$ and $\tilde{\omega}_{\mathrm{att} \pm}$ and then solve the resulting transcendental equation.

We apply this general scheme \cite{LikharevBook, Steiner2023.PhysRevLett.130.177002} to our minimal model at $A < 1/4$ and $\tilde{\phi} = \pi/2$ for the overdamped and underdamped regimes. Since we consider the low-temperature limit, the switching current is close to the critical one. We therefore use the asymptotic expressions \eqref{eq:Asympt U} and \eqref{eq:Asympt omega} for the barrier heights and oscillation frequencies, obtaining 
\begin{equation}
j_{\mathrm{sw} \pm} = j_{c \pm} -  
\begin{cases}
      \left(\frac{\theta}{u_{c\pm}}\ln \frac{2 \theta \omega_{c\pm}^2}{(6  \pi \ln 2) \varepsilon a u_{c\pm}} \right)^{2/3}, &
      \varepsilon \gg 1,\\ 
    \left(\frac{\theta}{u_{c\pm}}\ln \frac{2 \theta \omega_{c\pm}}{(6 \pi \ln 2) a  u_{c\pm}} \right)^{2/3}, &
    \varepsilon \ll 1,
\end{cases} 
\end{equation}
where $j_{c\pm}$ is given by Eq.\ \eqref{eq: jc}.
Note that the expression for the switching current at $\varepsilon \ll 1$ is obtained with logarithmic accuracy, while at $\varepsilon \gg 1$ the number under the logarithm is exact.

As expected, in the overdamped regime the difference between the switching and critical currents is smaller than the same quantity in the underdamped regime (due to the factor $\varepsilon$ in the denominator). For completeness, we note that, as shown in Ref.\ \cite{Steiner2023.PhysRevLett.130.177002}, in the extremely underdamped regime, one finds $j_{c \pm} -j_{\mathrm{sw \pm}} \sim  (\theta \ln \varepsilon S_{\pm})^{2/3}$. This asymptotic behavior replaces our result $j_{c \pm} - j_{\mathrm{sw \pm}} \sim  (\theta \ln\theta)^{2/3}$ that was obtained in the underdamped regime.

\section{Discussion}
\label{sec:discussion}

The minimal model of a Josephson element demonstrating the JDE is given by Eq.\ \eqref{eq:CPR}. It can be realized, e.g., in asymmetric higher-harmonic SQUIDs. The higher harmonics of the single-junction CPR naturally arise in various types of JJs with not too low transparencies of their weak-link regions \cite{LikharevBook,Golubov2004review}. At the same time, we note that effective CPR of this and actually arbitrary form can be engineered with the help of purely sinusoidal JJs connected in series and possibly in multiloop configurations \cite{Haenel2022arXiv,Gupta2023,Bozkurt2023.10.21468/SciPostPhys.15.5.204,Frattini2017.10.1063/1.4984142}.

\begin{figure*}[t]
\includegraphics[width=0.66\columnwidth]{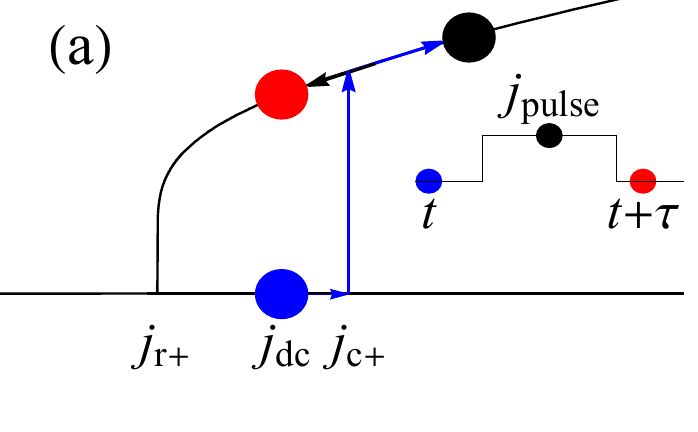}
 \hfill
  \includegraphics[width=0.66\columnwidth]{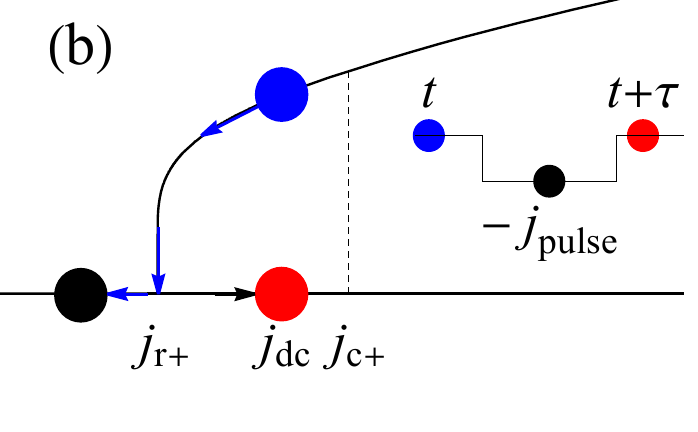} \hfill
   \includegraphics[width=0.66\columnwidth]{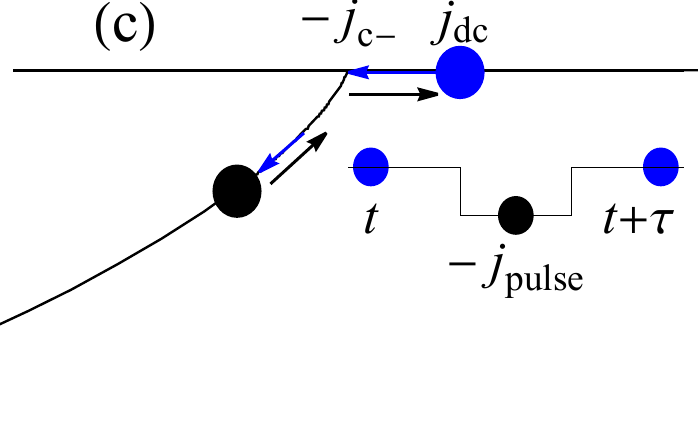}
\caption{Control of the diode state by applying the rectangular current pulse with amplitude $j_{\mathrm{pulse}}$ and duration $\tau \gtrsim \omega^{-1}_{J}$ in the regime of single-sided hystersis:
(a) and (b)~hysteretic CVC for positive current direction, and (c)~nonhysteretic CVC for negative current direction. Colored points represent different junction states at different moments in time: at the initial moment $t$ (blue point), when the pulse is applied (black point), and after the end of the pulse at $t + \tau$ (red point if the final state differs from the initial one and blue point otherwise). (a)~Evolution of the junction from the initial S state. Under the pulse action, the junction switches to the R state if $j_{\mathrm{dc}} + j_{\mathrm{pulse}} >j_{c+}$. In this case, when the pulse ends, the junction final state is on the R branch. (b)~Evolution of the junction from the initial R state. Under the pulse action, the junction switches to the S state if $j_{\mathrm{dc}} - j_{\mathrm{pulse}} < j_{r+}$. In this case, when the pulse ends, the junction final state is on the S branch.  (c)~Evolution of the junction from the initial S state in the nonhysteretic regime. After the pulse action, the junction returns to the initial state.}
 \label{fig: Oper of SSH}
\end{figure*}

As discussed in Sec.~\ref{sec:irradiation}, when the system is exposed to small external ac current, the Josephson plasma resonances in junctions with $\varepsilon \ll 1$ arise when the  external frequency satisfies the resonance condition $\tilde{\omega} \simeq \omega_{A \pm}(j_{\mathrm{dc}})$. In this case, the ac current can greatly enhance the thermal escape rate for one current direction (which satisfies the resonance condition) and thus stimulate switching from the S to R state, while leaving the system in the S state for the opposite current direction (which does not satisfy the resonance condition). As shown in Fig.~\ref{fig:Omega}, this behavior occurs at  $|j_{\mathrm{dc}}| < j_{c\pm}$. Therefore, this makes it possible to expand the operational range of the Josephson diodes when external ac current is applied. To this end, it is preferable to apply frequency $\tilde{\omega}$ in resonance with the largest of the two values $\omega_{A \pm}$. This is due to the fact that the enhancement factor of thermal escapes $\gamma (\tilde{\omega})$ rapidly decreases at $\tilde{\omega} > \omega_{A}(j_{\mathrm{dc}})$ \cite{Larkin1986, Devoret1987.PhysRevB.36.58, Krasnov2024.PhysRevApplied.22.024015}. The above procedure helps avoid parasitic switching from the S to R state in the opposite current direction.

The single-sided hysteresis also provides new opportunities for tuning the JDE, as illustrated in Fig.~\ref{fig: Oper of SSH}. Assume that the negative part of the CVC has only one stable branch at a fixed current value: the S state at $|j_{\mathrm{dc}}| < j_{c-}$ and the R state at $|j_{\mathrm{dc}}| > j_{c-}$, while the positive part of the CVC has two stable branches in the range $j_{r+}<j_{\mathrm{dc}} < j_{c+}$. In the latter case, it is possible to switch the system from the S to R state (or backward from the R to S state) by applying the rectangular current pulse with appropriate amplitude $j_{\mathrm{pulse}}$ and with pulse duration $\tau \gtrsim \omega_{J}^{-1}$. The amplitude $j_{\mathrm{pulse}}$ should satisfy the condition $j_{\mathrm{dc}} + j_{\mathrm{pulse}} > j_{c+}$ for switching from the S to R state [see Fig.~\ref{fig: Oper of SSH}(a)] and the condition $j_{\mathrm{dc}} - j_{\mathrm{pulse}} < j_{r+}$ for switching from the R to S state [see Fig.~\ref{fig: Oper of SSH}(b)]. At the same time, this pulse does not change the junction state in the case of negative $j_{\mathrm{dc}}$ [due to the absence of hysteresis in this direction, see Fig.~\ref{fig: Oper of SSH}(c)]. As a result, the above procedure makes it possible to change the diode state in one current direction while leaving the diode state intact in the opposite current direction.

Overall, our consideration demonstrates new possibilities for Josephson diode control in the case of finite capacitance of the junctions. For manipulation of the diode state by resonant ac current, junctions with $\beta \gg 1$ are preferable. At the same time, junctions with $\beta \sim 1$ open up additional ways of control in the regime of single-sided hysteresis while still being protected from strong suppression of the JDE by capacitance in the R state (generally, asymmetry of the CVC in the nonstationary regime weakens as $\beta$ increases). Finally, in the context of thermal fluctuations, junctions with $\beta \ll 1$ could be more practical because they are more stable with respect to temperature. For example, due to $j_{r} = j_{c}$ it is easier to return them to the S state after a thermal escape. 
 
\section{Conclusions}
\label{sec:conclusions}

In the framework of the RCSJ model, we have theoretically investigated the influence of junction capacitance and thermal fluctuations on the JDE in asymmetric higher-harmonic SQUIDs. In this model, the strength of charging and temperature effects is determined by the McCumber parameter $\beta$ and dimensionless temperature $\theta$.
In our work, we mainly focused on the minimal model in which the CPR of the SQUID in addition to the first Josephson harmonic also has the second one with dimensionless amplitude $A$ and phase shift $\tilde{\phi}$, see Eq.\ \eqref{eq:CPR}. We employed a combination of various perturbative methods, explicit analytical calculations, and numerical analysis to describe asymmetries of the CVC. 
Efficiency of the JDE and its polarity are determined by $\tilde\phi$ and
thus depend on the external magnetic flux $\Phi$.

In the presence of nonzero capacitance $\beta \neq 0 $, the CVC of the system may become hysteric and consist of two branches corresponding to the R and S states. Two new qualitative features arise in this case. One of them is asymmetry of the retrapping currents $j_{r\pm}$ and the second one is the single-sided hysteresis which can be observed within a certain range of $\beta$, see Fig.~\ref{fig:single sided hyst}. In this range, the system demonstrates qualitatively different behavior for different current directions (hysteretic CVC in one direction and  nonhysteretic CVC in the opposite one).

The oscillation frequency in the S state $\omega_{A \pm}$ of such a device depends on the current direction. This leads to asymmetric resonances and correspondingly to different values of the switching currents $j_{\mathrm{sw}\pm}$ in the presence of external ac irradiation.   

At the same time, the JDE is suppressed in the R state with increasing junction capacitance. Particularly, in the presence of ac irradiation this phenomenon manifests itself in weakening of asymmetry of the Shapiro steps as $\beta$ grows. 

Thermal fluctuations at $\theta \neq 0 $ lead to modifications of the CVC due to thermal activation processes. At $\beta = 0$, in the low-temperature limit we implemented the Ambegaokar-Halperin method and obtained exponentially strong asymmetry of the CVC analytically for arbitrary CPR at currents below the critical values. At $\beta \neq 0$, we calculated the asymmetric lifetimes $\tilde{\tau}_{l\pm}$ of the S states and then obtained expressions for the thermal switching currents $j_{\mathrm{sw} \pm}$.

 \acknowledgments
We thank V.~S.\ Stolyarov for useful discussions.
The work was supported by the Russian Science Foundation (Grant No.\ 24-12-00357).


\appendix

\section{Harmonic perturbation theory}
\label{Appendix:HarmPertTheory}

Below, we demonstrate how the HPT works in calculations of the corrections to the CVC in the large-capacitance limit and of the heights of the first Shapiro steps in the small-capacitance limit.

\subsection{CVC in the large-capacitance limit}
\label{sec: Appendix A CVC}

To apply the HPT in the large-capacitance limit, we represent the phase and voltage as series, see Eqs.\ \eqref{eq:phi as series} and \eqref{eq:series for v}.
Then, we substitute these expansions to Eq.\ \eqref{eq: Josephson equation without jac and noise}, expand the equation into the Fourier series, and solve it in the required order of the perturbation theory assuming conditions \eqref{eq:large capacitance limit}.

\subsubsection{First order}

In the first order, we need to take into account only the first and second harmonics in the Fourier series. The expansions  \eqref{eq:phi as series} and \eqref{eq:series for v} then take the following form:
\begin{widetext}
\begin{equation}
\varphi^{(1)}(\tau) = \overline{v}^{(1)} \tau + a^{(1)}_{1} \cos \overline{v} \tau + a^{(1)}_{2} \cos 2 \overline{v} \tau + b^{(1)}_{1} \sin \overline{v} \tau + b^{(1)}_{2} \sin 2 \overline{v}\tau, \qquad\overline{v} = j_{\mathrm{dc}} + \overline{v}^{(1)}. 
\end{equation}
In the leading order of the HPT,
\begin{equation}
-\beta j_{\mathrm{dc}}^2 (a^{(1)}_{1} \cos \overline{v} \tau + 4 a^{(1)}_{2} \cos 2 \overline{v} \tau +  b^{(1)}_{1} \sin \overline{v} \tau + 4 b^{(1)}_{2}\sin 2 \overline{v}\tau) = -\overline{v}^{(1)} - \sin \overline{v} \tau - A \sin (2 \overline{v} \tau - \tilde{\phi}).
\end{equation}
Note that the condition $\beta j_{\mathrm{dc}} \gg 1$ allows us to neglect the contributions arising from the dissipative term in Eq.\ \eqref{eq: Josephson equation without jac and noise} in this order of the HPT.

As a result, we obtain
\begin{align}
\overline{v}^{(1)}=0, \quad a_{1}^{(1)} = 0, \quad b_{1}^{(1)} = 1/\beta j_{\mathrm{dc}}^2, \quad a_{2}^{(1)} = -A \sin (\tilde{\phi})/4 \beta j_{\mathrm{dc}}^2, \quad b_{2}^{(1)} = A \cos(\tilde{\phi})/4\beta j_{\mathrm{dc}}^2.
\end{align}
In this order, there is no correction to the average voltage.

\subsubsection{Second order}

In the second order, we need to take into account the corrections to the first and second harmonics in the Fourier series, arising from the dissipative term, and also correction to the average voltage arising from the supercurrent term in Eq.\ \eqref{eq: Josephson equation without jac and noise}: 
\begin{equation}
\varphi^{(2)}(\tau) =  \overline{v}^{(2)} \tau + a^{(2)}_{1} \cos \overline{v} \tau + a^{(2)}_{2} \cos 2 \overline{v} \tau + b^{(2)}_{1} \sin \overline{v} \tau + b^{(2)}_{2} \sin 2 \overline{v} \tau ,\quad \overline{v} = j_{\mathrm{dc}} + \overline{v}^{(2)}. 
\end{equation}
Equation \eqref{eq: Josephson equation without jac and noise} then takes the following form: 
\begin{align}
&-\beta j_{\mathrm{dc}}^2 (a^{(2)}_{1} \cos \overline{v} \tau + 4 a^{(2)}_{2} \cos 2 \overline{v} \tau + 9 a^{(2)}_{3} \cos 3\overline{v} \tau + 16 a^{(2)}_{4} \cos 4 \overline{v} \tau +   b^{(2)}_{1}\sin \overline{v} \tau + 4 b^{(2)}_{2}\sin 2 \overline{v}\tau +  9 b^{(2)}_{3}\sin 3\overline{v} \tau + 16 b^{(2)}_{2}\sin 4 \overline{v}\tau) \notag
\\
&=-j_{\mathrm{dc}}(-a^{(1)}_{1} \sin \overline{v} \tau - 2 a^{(1)}_{2} \sin 2 \overline{v} \tau +  b^{(1)}_{1}\cos \overline{v} \tau + 2 b^{(1)}_{2}\cos 2 \overline{v}\tau) -\overline{v}^{(2)}  - (1/2) [a_{1}^{(1)} +  2 A  a_{2}^{(1)}\cos \tilde{\phi}  + 2 A  b_{2}^{(1)} \sin \tilde{\phi}].
\end{align}
Note that the constant (nonoscillating) corrections in Eq.\ \eqref{eq: Josephson equation without jac and noise} arise from  expansion of $\sin \varphi + A \sin(2 \varphi - \tilde{\phi})$. These terms are responsible for corrections to the average voltage (in this order, it is the last term in the square brackets). As a result,
\begin{align}
a_{1}^{(2)} = \frac{1}{\beta^2 j_\mathrm{dc}^3}, \quad b_{1}^{(2)} = 0, \quad a_{2}^{(2)} = \frac{A \cos \tilde{\phi}}{8 \beta^2 j_{\mathrm{dc}}^3}, \quad b_{2}^{(2)}= \frac{A \sin \tilde{\phi}}{8 \beta^2 j_{\mathrm{dc}}^3}, \quad \overline{v}^{(2)} = -\frac{a_{1}^{(1)} + 2 A (a_{2}^{(1)} \cos \tilde{\phi}  + b_{2}^{(1)} \sin \tilde{\phi} )}{2 \beta j_{\mathrm{dc}}^2}  = 0.
\end{align}
In this order, the corrections to the average voltage cancel each other.

\subsubsection{Higher orders}

Continuation of the above procedure to higher orders of the HPT is straightforward. One should substitute the expansion \eqref{eq:phi as series} into Eq.\ \eqref{eq: Josephson equation without jac and noise} in each order of the perturbation theory, solve the resulting equation for the Fourier coefficients, and then collect the constant terms [which appear from the expansion of $\sin \varphi + A \sin(2 \varphi - \tilde{\phi})$] which determine the correction to the average voltage. As a result of this procedure, we obtain the asymmetric CVC \eqref{eq:CVC in large capacity limit}.

\subsection{First Shapiro steps}
\label{sec: Appendix A Shapiro}

We also employ the HPT in the presence of the ac current to calculate asymmetry of the heights of the first Shapiro steps $\overline{v} = \pm\omega$. To this end, as mentioned in the main text, we slightly modify the HPT. We fix the average voltage  $\overline{v} = \mathrm{const}$ and find the corresponding current $j_{\mathrm{dc}}(\overline{v})$. Technically, we substitute the expansion \eqref{eq:phi as series ac} into Eq.\ \eqref{eq: J1 with ac} and then solve the equation in the required order of the perturbation theory. As mentioned in Sec.~\ref{sec:irradiation}, the HPT works in the two limiting cases, the large- and small-capacitance limit [defined by Eqs.\ \eqref{eq:large capacitance limit} and \eqref{eq:small capacitance limit}, respectively].

To demonstrate how the HPT works in this case, we calculate the heights of the first Shapiro steps at $\beta = 0$. We emphasize that at $\beta \neq 0$, the leading asymmetric term in the heights will be the same. Similarly, the results for the large-capacitance limit can be obtained by this technique.

\subsubsection{HPT for the first Shapiro steps at
\texorpdfstring{$\beta = 0$}{beta = 0}}

At $\beta = 0$, the first Josephson equation \eqref{eq: J1 with ac} takes the following form:
\begin{equation}
 \dot{\varphi} + \sin\varphi + A \sin(2 \varphi - \tilde{\phi}) = j_{\mathrm{dc}} + j_{\mathrm{ac}} \cos(\omega \tau + \delta).
\end{equation}
In the first order of the perturbation theory, we write
\begin{equation}
\varphi^{(1)}(\tau) = a^{(1)}_{1}\cos \overline{v}\tau + b^{(1)}_{1}\sin \overline{v}\tau + a^{(1)}_{2}\cos 2\overline{v}\tau + b^{(1)}_{2}\sin 2\overline{v}\tau, \qquad
j_{\mathrm{dc}} = \overline{v} + j^{(1)}.
\end{equation}
Solving the resulting equation on the Fourier coefficients,
\begin{equation}
 \overline{v}(-a^{(1)}_{1}\sin \overline{v}\tau + b^{(1)}_{1}\cos \overline{v}\tau - 2a^{(1)}_{2}\sin 2\overline{v}\tau + 2b^{(1)}_{2}\cos 2\overline{v}\tau)  = j^{(1)} + j_{\mathrm{ac}} \cos(\overline{v} \tau + \delta) - \sin \overline{v} \tau - A \sin(2\overline{v} \tau - \tilde{\phi}),
 \end{equation} 
we find the solution 
\begin{equation}
\varphi^{(1)} = \left[j_{\mathrm{ac}}\sin(\overline{v} \tau + \delta) + \cos \overline{v} \tau + (A/2) \cos (2 \overline{v} \tau - \tilde{\phi} )\right]/\overline{v}, \qquad j^{(1)} = 0.
\end{equation}

In the second order of the HPT, we write
\begin{equation}
\dot \varphi^{(2)} = j^{(2)} + \varphi^{(1)}(\tau) \left[ \cos \overline{v} \tau + 2 A \cos(2 \overline{v} \tau - \tilde{\phi}) \right].
\end{equation}
The solution of this equation takes the form
\begin{align} \label{eq:phi 2 Shapiro step}
 \varphi^{(2)}= \frac{1}{24 \overline{v}^2}\Bigl[-30 A \sin (\overline{v} \tau-\tilde{\phi}) -24 A j_\mathrm{ac} \cos (\overline{v} \tau - \tilde{\phi} - \delta ) &+6 j_\mathrm{ac} \cos (2 \overline{v} \tau + \delta )-6 \sin (2 \overline{v} \tau) \notag  \\ + 8 A j_\mathrm{ac} \cos (3 \overline{v} \tau + \delta - \tilde{\phi}) & - 3 A^2 \sin (4\overline{v} \tau -2\tilde{\phi})- 10 A \sin (3 \overline{v} \tau -\tilde{\phi})\Bigl], 
 \\ 
 \label{eq:j^2 HPT}
 j^{(2)} =  (1  + A^2 &+ j_{\mathrm{ac}} \sin\delta)/2 \overline{v}.
\end{align}
\end{widetext}
The first two terms in the expression \eqref{eq:j^2 HPT} produce corrections to Ohm's law, and they are present even at $j_{\mathrm{ac}} = 0$. The last term depends on $j_{\mathrm{ac}}$ and is responsible for the heights of the Shapiro steps. As mentioned earlier, the initial phase $\delta$ can be arbitrary. The last term in Eq.\ \eqref{eq:j^2 HPT} can thus take different values depending on $\delta$. As a result, one particular voltage $\overline{v} = \pm \omega$ corresponds to a range of current values $j_{\mathrm{dc}}$ (hence, the step in the CVC). In the lowest order of the perturbation theory, the heights of the first Shapiro steps are equal to $j_{\mathrm{ac}}/\omega$. 

Continuing this procedure in the next order of the perturbation theory, we obtain
\begin{equation} \label{eq:j^3 HPT}
j^{(3)} = - \frac{3 A \sin \tilde{\phi}}{4 \overline{v}^2}-\frac{9 A j_{\mathrm{ac}} \cos (\tilde{\phi} + \delta)}{8 \overline{v}^2}. 
\end{equation}
Note that both the corrections \eqref{eq:j^2 HPT} and \eqref{eq:j^3 HPT} at $j_{\mathrm{ac}} =0$ coincide with the correction to the CVC obtained in Ref.\ \cite{Fominov2022.PhysRevB.106.134514}. 
The ac-dependent correction to the current can be written in the following form:
\begin{equation}
\Delta j = \frac{j_{\mathrm{ac}}}{2 \omega}\left[\left(1 \pm \frac{9 A \sin \tilde{\phi}}{4 \omega} \right) \sin\delta \mp \frac{9 A \cos \tilde{\phi}}{4 \omega} \cos\delta \right].
\end{equation}
As a result, in the leading order of the HPT, asymmetry of the heights of the first Shapiro steps takes form of Eq.\ \eqref{eq:Shapiro steps in small capacitance limit}.
Note that in this equation, $j_{\mathrm{ac}}$ and $A$ are assumed to be small only compared to large $\omega$.

\section{Calculation of the retrapping currents}
\label{Appendix:Retrap}

Below, we calculate the retrapping currents $j_{r\pm}$ employing the perturbation theory with respect to the small parameters $A \ll 1$ and $\varepsilon \ll 1$. We apply the general scheme  described in Sec.~\ref{sec:Retrapping current}, sequentially in each step of the perturbation theory. For convenience, we assume $j_{\mathrm{dc}} > 0$ and find $j_{r+}$. To obtain $j_{r-}$, we only need to substitute $\tilde{\phi} \mapsto - \tilde{\phi}$ in the final expression.

\subsection{First order with respect to \texorpdfstring{$\varepsilon$}{varepsilon} and the zeroth order with respect to \texorpdfstring{$A$}{A}}

We start by reproducing the well-known answer for the retrapping current in the large-capacitance limit at $A = 0$ (without the JDE). 
The expressions for $\varphi_{\max}$ and $E_{0}$ take the form
\begin{equation}
\varphi_{\max} = -\pi, \quad E_0 = 1.
\end{equation}
Due to weak dissipation ($\varepsilon \ll 1$), in this order of the perturbation theory we can neglect both the dissipative term in Eq.\ \eqref{eq:E(t)} and $j_{r}$ in $U(\varphi)$. Additionally, we neglect the $A$ term in the potential. As a result, we obtain
\begin{equation} \label{eq:dotvarphi in 0 order}
 \dot{\varphi} =  \sqrt{2 (1+ \cos\varphi)}.
\end{equation}
From Eq.\ \eqref{eq:j_{r}}, we then find
\begin{equation} \label{eq:jr in 0 order}
   j_{r_{\pm}} = \frac{\varepsilon}{2 \pi}  \int^{\pi}_{- \pi}  
    \sqrt{2(1 + \cos\varphi)} d \varphi = \frac{4 \varepsilon}{\pi}.
\end{equation}

\subsection{First order with respect to \texorpdfstring{$\varepsilon$}{varepsilon} and to \texorpdfstring{$A$}{A}}

Next, we consider the effect of the second harmonic on the retrapping current in the first order of the perturbation theory with respect to $A$. Corrections to $\varphi_{\max}$ and $E_{0}$ take the following form:
\begin{equation}
 \varphi_{\max} = -\pi - A \sin\tilde{\phi},\quad E_{0} = 1 - (A/2) \cos\tilde{\phi}.
\end{equation}
In this order of the perturbation theory, we must take into account the $A$ term in $U(\varphi)$ but can still ignore $j_{r\pm}$ and the dissipative term in Eq.\ \eqref{eq:E(t)}. As a result,
\begin{gather}
    \dot{\varphi} =\sqrt{2(1 + \cos\varphi) + A[\cos(2 \varphi - \tilde{\phi}) - \cos\tilde{\phi}]}, \label{eq:dotphi in 1 order} \\
   j_{r\pm}  = (4 \varepsilon/\pi) \bigl[1 - A \cos(\tilde{\phi})/3 \bigr]. 
\label{eq:jr in the first order in A}
\end{gather}
Equation\ \eqref{eq:jr in the first order in A} takes into account the corrections to the retrapping current in the first order with respect to $A$. However, the retrapping current in this order is symmetric and, as one can check, this will be so in any order with respect to $A$ in the first order with respect to $\varepsilon$ [because in the first order of the perturbation theory with respect to $\varepsilon$ we neglect the dc-current contribution in Eq.\ \eqref{eq:E(t)}]. Therefore, we need to consider the next order of the perturbation theory to find asymmetry of the retrapping currents $j_{r+} \neq j_{r-}$.

\subsection{Second order with respect to \texorpdfstring{$\varepsilon$}{varepsilon} and the first order with respect to \texorpdfstring{$A$}{A}}

In this order, we replace $j_{\mathrm{dc}}$ by its value from the previous step of the perturbation theory [Eq.\ \eqref{eq:jr in the first order in A}] and $\dot{\varphi}$ in the dissipative term in the rhs of Eq.\ \eqref{eq:E(t)} with the value from Eq.\ \eqref{eq:dotphi in 1 order}. However, first we determine the location of the potential maximum and the initial energy:
\begin{align}
\varphi_{\max} = -\pi &- A \sin \tilde{\phi}- \frac{4 \varepsilon}{\pi}\left(1 + \frac{5  A \cos\tilde{\phi}}{3}\right), \label{eq: phi max}
\notag \\
E_{0} = 1 - \frac{A \cos \tilde{\phi}}{2} 
&+ \frac{4 \varepsilon  A  \sin \tilde{\phi}}{\pi}+ 4 \varepsilon \left({1- \frac{A \cos\tilde{\phi}}{3}}\right).
\end{align}
After that, we calculate corrections to $\dot{\varphi}$ from Eq.\ \eqref{eq:E(t)} with $\varphi_{\mathrm{in}} = \varphi_{\max}$:
\begin{widetext}
\begin{multline} \label{eq: dot phi A and varepsilon}
 \dot{\varphi} = \Biggl[2\left(1 + \cos\varphi\right) + A\left(\cos (2 \varphi - \tilde{\phi}) - \cos\tilde{\phi} \right)  +  \frac{8 \varepsilon (\varphi + \pi) }{\pi}\left(1 - \frac{A \cos \tilde{\phi}}{3}\right)   
 \\
 -2 \varepsilon \int^{\varphi}_{-\pi - A \sin\tilde{\phi}} \sqrt{2(1 + \cos x) + A\left(\cos(2 x - \tilde{\phi}) - \cos\tilde{\phi} \right)} d x \Biggr]^{1/2}.
\end{multline}
We substitute expression \eqref{eq: dot phi A and varepsilon} for $\dot{\varphi}$ and expression \eqref{eq: phi max} for $\varphi_{\max}$ to Eq.\ \eqref{eq:j_{r}}.
After that, we expand the result to the first order with respect to $A$ and $\varepsilon$, and calculate the integrals. As a result, we obtain Eq.\ \eqref{eq:ans j_{r}}.

\section{Calculation of the CVC at \texorpdfstring{$T \neq 0$}{T != 0} in the zero-capacitance limit}
\label{sec:Ambegaokar for min model}

We use the general formula \eqref{eq:Zcalcul} assuming that $\theta \ll A \ll 1$ and keeping only the leading terms with respect to $A$. We also assume that $j_{\mathrm{dc}}>0$ and use the general symmetry \eqref{eq:symmetry} to obtain the negative branch of the CVC. In this limit, the potential has only one minimum and maximum per period. Their locations are given by
\begin{align}
\varphi_{\min \pm} = \arcsin j_{\mathrm{dc}} -2 A j_{\mathrm{dc}} \cos \tilde{\phi} \mp\frac{2 A j_{\mathrm{dc}}^2 \sin \tilde{\phi}}{\sqrt{1-j_{\mathrm{dc}}^2}} \pm \frac{A \sin \tilde{\phi}}{\sqrt{1-j_{\mathrm{dc}}^2}},
\\
\varphi_{\max\pm, m} = 2 \pi m +\pi - \arcsin j_{\mathrm{dc}} -2 A j_{\mathrm{dc}} \cos \tilde{\phi} \pm \frac{2 A j_{\mathrm{dc}}^2 \sin \tilde{\phi}}{\sqrt{1-j_{\mathrm{dc}}^2}} \mp \frac{A \sin \tilde{\phi}}{\sqrt{1-j_{\mathrm{dc}}^2}}.
\end{align}
The values of the potential energy and its second derivative at the extrema are given by expressions
\begin{align}
U_{\mathrm{asym\pm}}(\varphi_{\min\pm}) = -\sqrt{1-j_{\mathrm{dc}}^2}\left(1 \pm A j_{\mathrm{dc}} \sin \tilde{\phi} \right)-j_{\mathrm{dc}} \arcsin j_{\mathrm{dc}}  
&+ A \left(j_{\mathrm{dc}}^2  - 1/2 \right) \cos \tilde{\phi}  ,
\\
U_{\mathrm{asym\pm}}(\varphi_{\max\pm,m}) = \sqrt{1-j_{\mathrm{dc}}^2}\left(1 \pm A j_{\mathrm{dc}} \sin \tilde{\phi} \right)-\pi  j_{\mathrm{dc}}(2 m +1) &+j_{\mathrm{dc}}\arcsin j_{\mathrm{dc}}  +A \left(j_{\mathrm{dc}}^2  - 1/2 \right) \cos \tilde{\phi} ,
\\
U_{\mathrm{asym\pm}}''(\varphi_{\min\pm})=\sqrt{1-j_{\mathrm{dc}}^2}\left(1 \pm A j_{\mathrm{dc}}\frac{ 3 - 2j_{\mathrm{dc}}^2}{1 - j_{\mathrm{dc}}^2} \sin \tilde{\phi}\right) &+ 2A(1 - j_{\mathrm{dc}}^2)\cos\tilde{\phi},
\\
|U_{\mathrm{asym\pm}}''(\varphi_{\max\pm, m})|=\sqrt{1-j_{\mathrm{dc}}^2}\left(1 \pm A j_{\mathrm{dc}}\frac{ 3 - 2j_{\mathrm{dc}}^2}{1 - j_{\mathrm{dc}}^2} \sin\tilde{\phi}\right) &- 2A(1 - j_{\mathrm{dc}}^2)\cos\tilde{\phi}.
\end{align}
The sum over $m$ in Eq.\ \eqref{eq:Zcalcul} can be easily calculated as a sum of a geometric progression and yields the factor $\left[1 - \exp{(- 2 \pi j_{\mathrm{dc}}/\theta)}\right]^{-1}$.

Substituting these expressions into Eqs.\ \eqref{eq:Zcalcul} and \eqref{eq:fluctV}, we obtain Eq.\ \eqref{eq:answer Ambegaokar} for the CVC.

\end{widetext}



%

\end{document}